\documentclass[fleqn,usenatbib,onecolumn]{mnras}

\usepackage{newtxtext,newtxmath}

\usepackage[T1]{fontenc}
\usepackage{ae,aecompl}

\usepackage{graphicx}
\usepackage{epstopdf}
\usepackage{deluxetable}
\usepackage{amsmath}
\usepackage{natbib}

\newcommand{\chisq}{$\chi^2$}

\newcommand{\ecosw}{$e\cos\omega$}
\newcommand{\esinw}{$e\sin\omega$}
\newcommand{\kepler}{\emph{Kepler}\ }
\newcommand{\rsun}{$R_{\odot}$}
\newcommand{\msun}{$M_{\odot}$}
\newcommand{\edit}[1]{{#1}}

\title[Modeling Kepler Eclipsing Binaries]{Modeling \kepler Eclipsing Binaries: Homogeneous Inference of Orbital \& Stellar Properties}

\author[D. Windemuth et al.]{
D. Windemuth,$^{1}$\thanks{E-mail: windemut@uw.edu}
E. Agol,$^{1, 2}$
A. Ali,$^{1}$
F. Kiefer$^{3,4}$
\\
$^{1}$Department of Astronomy, University of Washington, Seattle WA 98195, USA\\
$^{2}$Guggenheim Fellow\\
$^{3}$Institut d`Astrophysique de Paris, 75014 Paris, France\\
$^{4}$CNES Fellow\\
}
\date{Accepted 2019 July 30. Received 2019 July 26; in original form 2018 August 17}

\pubyear{2019}

\begin{document}

\label{firstpage}
\pagerange{\pageref{firstpage}--\pageref{lastpage}}
\maketitle

\begin{abstract}
We report on the properties of eclipsing binaries from the \kepler mission with a newly developed photometric modeling code, which uses the light curve, spectral energy distribution of each binary, and stellar evolution models to infer stellar masses without the need for radial velocity measurements. We present solutions and posteriors to orbital and stellar parameters for 728 systems, forming the largest homogeneous catalogue of full \kepler binary parameter estimates to date. Using comparisons to published radial velocity measurements, we demonstrate that the inferred properties (e.g., masses) are reliable for well-detached main-sequence binaries, which make up the majority of our sample. The fidelity of our inferred parameters degrades for a subset of systems not well described by input isochrones, such as short-period binaries that have undergone interactions, or binaries with post-main sequence components.  Additionally, we identify 35 new systems which show evidence of eclipse timing variations, perhaps from apsidal motion due to binary tides or tertiary companions. We plan to subsequently use these models to search for and constrain the presence of circumbinary planets in \kepler eclipsing binary systems.
\end{abstract}

\section{Introduction}

A significant fraction of stars live in binaries or higher-order multiple systems \citep{Abt1979, Duchene2013}; despite this, the effects of binarity on a system's formation, stellar evolution, and dynamical evolution, including planetary system evolution, are not well understood. These uncertainties propagate to other fields of astronomy. For example, they have profound implications on interpretations of stellar populations, the foundation of galactic and extragalactic astronomy \citep{Eldridge2017,Izzard2018}. Constraining binary processes requires knowledge of their fundamental parameters, such as radii, temperatures, mass ratio, and age. Fortunately, the geometry of eclipsing binary (EB) stars allows for direct constraints on such parameters. This makes EBs excellent astrophysical tools not only to calibrate single star stellar evolution models  \citep[e.g.,][]{Stassun2009, Claret2018, delBurgo2018}, but also to test models of binary tidal evolution \citep{Birkby2012,Lurie2017,Fleming2019}, along with their planetary systems \citep{Martin2015,Fleming2018}. However, direct measurements of binary absolute dimensions rely on the fortuitous condition that the binary is an eclipsing, double-lined spectroscopic system, and necessitate multi-band time-series and radial velocity (RV) information, which is observationally expensive to obtain for a large ensemble of systems. Making meaningful inferences about \emph{populations} of binaries, therefore, may benefit from a different approach.

The goal of this work is to estimate absolute stellar and orbital dimensions for a large ensemble of $\sim$detached eclipsing binaries in the \kepler field, \emph{relying only on photometry}. Rather than use time-intensive measurements of a small number of bright EBs to constrain stellar models, we leverage theoretical stellar isochrones to infer properties for a much larger sample of systems. The basis for our study begins with the \edit{Villanova \kepler EB Catalogue (VKEBC; \citealp{Prsa2011, Kirk2016})}, which contains over 2800 eclipsing and ellipsoidal binaries from the \kepler mission.  The majority of these objects have only crude phenomenological estimates of eccentricity, temperature ratios, and sum of radii relative to orbits, without a full characterization of the binary parameters. Detailed, full parameter studies have typically focused on individual systems or smaller samples of binaries for e.g., asteroseismic \citep{Guo2016, Gaulme2016}, eccentricity-tidal \citep{Kjurkchieva2016}, and RV analysis \citep{Matson2017}. Here, we statistically combine the constraining power of both light curves and SEDs to derive full system solutions and posteriors to 728 \kepler EBs, creating a large homogeneous catalogue.

Such a catalogue has yet to be developed for the binaries in the \kepler field, although comparable analyses have been carried out for other catalogues of eclipsing binaries. The most similar study was applied to the Trans-Atlantic Exoplanet Survey (TrES; a pre-cursor survey to \emph{Kepler}) by \citet{Devor2008} using light curves collected to discover transiting exoplanets from the ground, supplemented by multi-band photometry. The advantage of the \kepler field is that it forms a homogeneous dataset with high precision, continuous time-series photometry over a long baseline, and that it is extremely well studied in the characterization of single stars and exoplanets. 

The original impetus for our catalogue is to characterize a large population of binaries to a sufficient degree to enable a search for transiting circumbinary planets (CBPs) and to quantify the frequency of CBPs as a function of stellar binary properties (Windemuth et al., in prep). In addition to this, our analysis can provide the basis for additional studies of binary stellar populations.  Time-intensive radial velocity follow-up of \kepler EBs would benefit from informative priors on masses, radii, temperatures, and ages of a large EB sample to select the preferred targets to observe.  Our homogeneous parameter estimates could also permit more detailed statistical testing of physical processes, such as tidal circularization \citep{Duquennoy1991,VanEylen2016} and rotational synchronization \citep{Lurie2017}, and could facilitate detection of additional bodies in the system, such as tertiary stars  \citep[e.g.,][]{Borkovits2015,Borkovits2016}. Our methodology may be applied to current and upcoming surveys in the era of large-scale precision photometry such as OGLE, \emph{TESS}, and LSST, which have all-sky coverage, opening up an avenue for binary star galactic archaeology. 

With these goals in mind, here we present a Bayesian characterization of 728 well- to semi-detached EBs using \kepler time-series photometry and archival multi-band imaging, as well as \emph{Gaia} parallaxes when available; that is, we derive full system parameters with only photometry and astrometry, and no RV information. Simultaneous modeling of the light curve (LC) and spectral energy distribution (SED) provides joint constraints on EB orbital (e.g., period, eccentricity, argument of periastron, inclination) and stellar (e.g., mass, radius, temperature, age) properties, subject to the assumption that the two stars fall on a single isochrone. The paper is organized as follows. In section \ref{sec:methods}, we describe our sample selection and the simultaneous LC + SED model. We present the results of our analysis in section \ref{sec:results}, and discuss our results in the context of stellar evolution, systematic biases, previous EB radial velocity and population studies in section \ref{sec:discussion}. In section \ref{sec:conclusions}, we conclude and summarize our key findings. 

\section{Methods}
\label{sec:methods}

We developed a Python code, dubbed ``KEBLAT", to simultaneously fit the light curves and SEDs of 728 \kepler EBs
to obtain their orbital and stellar parameters. \edit{The open-source code is available on Github}\footnote{\texttt{https://github.com/savvytruffle/keblat/}}.  In this section we describe 
the target selection process (\S \ref{subsec:sample}), the
datasets gathered (\S \ref{subsec:data_acquisition}), the light curve modeling (\S \ref{subsec:lightcurve}), spectral energy distribution modeling (\S \ref{subsec:sed}), and joint LC+SED modeling of each binary (\S \ref{subsec:joint} and \ref{subsec:fitting}). To test the accuracy of our model inference, we also extract RVs for 3 EBs in our sample that have not been studied previously (\S \ref{subsec:flavien_rvs}); we add these to published EBs with RV mass solutions to compare to our inferred masses below (\S \ref{subsec:mass}).  Throughout the paper, we refer to properties of the primary and secondary components with subscripts 1 and 2, respectively, and define the primary component as the eclipsed star that exhibits the deeper light curve minimum. 

\subsection{Sample selection} 
\label{subsec:sample}

We selected our binary systems from the Villanova \kepler EB Catalogue \edit{or VKEBC} \citep{Prsa2011, Kirk2016}. This sample, totaling 2877 targets, has been compiled from the entire \kepler prime mission and includes binaries which are not eclipsing (e.g., ellipsoidal variables). We made selection cuts to reduce grazing (i.e., non-constraining) geometries and the effects of close binarity that are not included in our model (e.g., tidal distortions and Doppler boosting of the binary components). 

First, to \edit{diminish} the presence of background, grazing, or non- eclipsing binaries, 
we required primary and secondary eclipse depths to be $>$5\%  and $>$0.1\% 
of the normalized flux, respectively. Next, we excluded targets which exhibit strong distortions in the light curve by selecting for binaries with morphology parameter \texttt{morph} $<$ 0.6. This morphology parameter, provided by \edit{VKEBC}, is based on a locally linear embedding scheme that classifies light curve shape, such that a low value correlates with well-detached systems exhibiting clearly separated eclipses, and higher morphology value correlates with over-contact (OC) systems with sinusoidal variations \citep{Matijevic2012}. 

In total, we investigated 728 $\sim$detached eclipsing binary systems. The large reduction in the number of systems from the full \edit{VKEBC} is related to the fact that short-period binaries have the highest eclipse probability, but also have the largest tidal distortions; our sample emphasizes the longer period eclipsing binaries.

\subsection{Data Acquisition}
\label{subsec:data_acquisition}
\subsubsection{\kepler photometric time series data} \label{subsec:kepler_data}

We extracted light curves from \kepler long-cadence ($\sim$30 min) simple 
aperture photometry (SAP) based on Data Release 24, spanning $\sim$4 years or 17 quarters.  To correct for slight variations in instrumental sensitivity across quarterly rotations, whereby light from a target may fall upon different detector and pixels, we normalized the raw SAP flux by the quarterly median value.  To lessen
the impact of outliers in the data, we then de-weighted data points with quality 
flags $>$ 8 (signifying cosmic ray hits, space craft re-pointing, etc) by 
inflating their error bars 10-fold. We did not fit the entire \kepler light curve for our final analysis. Instead, we clipped the light curve around each eclipse with a window 1.5--2.0 times eclipse durations, which were initially taken from \edit{VKEBC}, and then iteratively refined during the optimization process. This window around each eclipse allowed us to fit a polynomial to the variable component of each binary which is not included in our physical model, while minimizing the model evaluation and computational time. 

\subsubsection{Spectral energy distribution data} \label{subsec:sed_data}

As we did not have radial velocity data for most of our targets, we required additional
constraints to infer the absolute parameters of each system.  We collected literature
photometric data on each target, and modeled these with stellar evolution models
(see \S \ref{subsec:sed}). We identified extant photometry for each target by cross-matching our sample against sources from the Howell-Everett \citep[HE,][]{Everett2012}, Sloan Digitial Sky Survey  \citep[SDSS,][]{Ahn2012},  Two-Micron All-sky Survey \citep[2MASS,][]{Skrutskie2006}, and Wide-field Infrared Survey Explorer \citep[WISE,][]{Cutri2014} catalogues on Vizier using a 2$^{\prime\prime}$ search cone. We gathered together the available archival photometry to construct SEDs, excluding targets which had fewer than five
photometric data points. 

Having multiple measurements spanning a broad wavelength range is crucial 
to leverage the shape of stellar SEDs to constrain the stellar properties. The $U$ 
and $B$ flux, in particular, distinguish young and hot stars from cooler, older 
stars. While the HE and 2MASS campaigns have nearly complete coverage of the 
\kepler field, the WISE footprint on our sample covers only $\sim$50\%, and the
SDSS footprint covers only $\sim$10\%. Therefore, we supplemented SDSS photometry 
with $g^\prime\, r^\prime\,  i^\prime\, z^\prime$ estimates from the \kepler Input 
Catalogue (KIC), which were 
designed to mimic SDSS bandpasses. The typical agreement between KIC and SDSS 
\emph{griz} measurements for our sample is of order $\sim$0.05 mag. Furthermore, we used the measured ephemerides to identify that about 10--15\% of the photometric observations had been taken during an eclipse; we inflated their error bars to 3 mags to diminish their influence during fitting.

\subsection{Light curve model} \label{subsec:lightcurve}

\edit{The LC module couples a Keplerian orbit solver to the analytic \cite{Mandel2002} transit model with a quadratic limb-darkening law to fit the observed light curve. The Keplerian orbital model determines the instantaneous relative positions of the binary stars, which are used, in conjunction with radius ratio and a limb darkening prescription, to compute the obscured-to-unobscured system flux. We apply a triangular parameterization of the quadratic limb darkening coefficients to allow uniform prior sampling \citep{Kipping2013}; these coefficients are allowed to float and are not explicitly coupled to model atmospheres.}

\edit{Using this model,} KEBLAT constructs a template light curve with high temporal fidelity at $\sim$1min, incrementally summed to produce a $\sim$30 min integration, i.e., the \kepler long-cadence exposure time. The code then uses this high fidelity template to linearly interpolate the flux at each observed cadence. This method speeds up the computation time and accounts for distortions in the light curve due to finite integration times \citep{Kipping2010}. For stellar eclipses, which are much higher signal-to-noise than exoplanet transits, this step is especially important; the effect of finite integration time is strongest during ingress and egress, which in turn can affect inference of eccentricity, inclination, radii, and limb darkening coefficients. We test this interpolation scheme against direct computation of the light curve and found that the maximum error is $\sim10^{-6}$, at least 2--3 orders of magnitude smaller than typical in-eclipse residuals due to dynamical and stellar variability. We apply quarterly contamination values from \kepler to model (varying) third-light contributions, and allow each quarter's contamination value to float during optimization. 

In total, the light curve model has 13+$N_{\mathrm{quarters}}$ free parameters describing stellar properties ($M_{\mathrm{sum}}$, $\frac{R_2}{R_1}$, $R_{\mathrm{sum}}$, $\frac{F_2}{F_1}$), orbital elements ($e\sin\omega$, $e\cos\omega$, $b$, P, t$_{\mathrm{PE}}$), limb darkening ($q_{11}, q_{12}, q_{21}, q_{22}$), a systematic error term  $\sigma_{\mathrm{sys}}$ to account for underestimated observational uncertainties, and crowding values for quarter $j$ ($c_j$). Note that we define the impact parameter with respect to the primary, such that $b = a\cos i/R_1$. Transformed parameters such as \esinw, \ecosw, and limb darkening coefficients have finite bounds. For non-transformed variables, we place broad bounds; see Table~\ref{tab:modelparameters} for a detailed summary of each parameter.

During each model evaluation, we simultaneously fit a second order local polynomial around each eclipse to marginalize the background continuum flux, which may be influenced by instrument variations and/or stellar activity. We used a window of 1.5 -- 2 times the primary eclipse duration to ensure our polynomial choice is appropriate, and found that a second order polynomial was typically sufficient. Because the eclipses are masked during polynomial marginalization, higher order polynomials may overfit the data and potentially introduce spurious signals.

\subsection{Stellar evolution and photometric models} \label{subsec:sed}

The SED model assumes a co-eval binary specified by age $\tau$, metallicity $z$, and masses $M_1$ and $M_2$. We utilize PARSEC \citep{Bressan2012} stellar evolution isochrones, a large grid of stellar models featuring properties such as luminosity, $\log g$, effective temperature, and flux in various bandpasses. We adopt a fine grid spacing, with $z_\mathrm{grid}$ = [0.001, 0.002, 0.004, 0.008, 0.01, 0.015, 0.02, 0.03, 0.04, 0.05, 0.06], $\Delta \tau_{\mathrm{grid}}$ = 0.02, and median $\Delta M_{\mathrm{grid}}$ = 0.006. For a specified set of mass, age, and metallicity, KEBLAT performs a bi-linear interpolation over the PARSEC grid to determine the predicted temperatures and absolute magnitudes in \emph{UBV griz JHK $W_1W_2$} bandpasses, i.e., the stellar SED. We calculate the effective radius of each star from the bolometric luminosity via the Stefan-Boltzmann law; these radii values are necessary to couple to the LC module. 

To fit the archival data, we sum the predicted SEDs of both stellar components, compute the distance modulus, and correct for dust extinction along the line of sight assuming an exponential dust distribution with scale-height $h_0 = 119$ pc \citep{Kruse2014}:
\begin{equation}
\mathrm{mag}_{\lambda, \mathrm{binary}} = \mathrm{Mag}_{\lambda, \mathrm{binary}} + 5 \log_{10}\left(\frac{d}{10}\right) + A_{\lambda} E(B-V) \left(1-\exp{\left(\frac{-d \sin b_G}{h_0}\right)}\right) \,
\end{equation}
where $b_G$ is a target's galactic latitude, $d$ is distance in pc, and $E(B-V)$ and $A_{\lambda}$ are reddening and band specific extinction computed from a Milky Way extinction law with $R=3.1$ \citep{Fitzpatrick1999}. The integrated absolute magnitude of the binary in a given bandpass is
\begin{equation}
\mathrm{Mag}_{\lambda, \mathrm{binary}} = \mathrm{Mag}_{\lambda, 1} - 2.5 \log_{10} \left(1+10^{-0.4(\mathrm{Mag}_{\lambda, 2}-\mathrm{Mag}_{\lambda, 1})}\right) \ .
\end{equation}
We use cross-matched \emph{Gaia} distances derived from \cite{Bailer-Jones2018} and \cite{Schlafly2011} dust maps results, when available, to place Gaussian priors on $d$ and $E(B-V)$ in our model. Accurate distances from parallax may be used to place better constraints on EB masses since mass correlates tightly with luminosity on the main sequence.  Well-calibrated eclipsing binaries can be used as standard candles to calibrate parallaxes \citep{Southworth2005}, and the converse should also be true for systems with well-constrained geometries \citep{Stassun2016}. In particular, accurate distances may better constrain masses for binaries with \edit{non-total or non-annular eclipses} by precisely determining total system luminosities. However, rather than using reported uncertainties as fixed $\sigma_d$, $\sigma_{\mathrm{E(B-V)}}$, we allow the widths of these Gaussian priors to float to tolerate inaccuracies in the dust map or \emph{Gaia} data due to source confusion or presence of tertiary companions, which has an occurrence rate of $\sim15-20$\% in the \kepler field \citep{Gies2012,Rappaport2013, Conroy2014, Orosz2015}. Frequent eclipses or nearby long-period binaries may also deleteriously affect the accuracy of \emph{Gaia} astrometry.

As with the LC model, the SED model fits for a systematic error, $\sigma_{\mathrm{SED}}$, to account for underestimated observational errors and uncertainties in the isochrone models. This systematic error also encapsulates any contaminant flux to the binary by additional components, as we do not fit for a third star given the limited number of available SED measurements. As a result, we expect this systematic term to be relatively large compared to reported observational uncertainties.  

In practice, we use the sum and ratio of masses ($M_{\mathrm{sum}}=M_1+M_2$, $Q = M_2/M_1$), instead of individual masses ($M_1$, $M_2$), to optimize the SED fits. The SED module has in total 10 possible free parameters describing the fundamental stellar properties ($M_{\mathrm{sum}}$, $Q$, age, metallicity), dust scaleheight $h_0$,  $\sigma_{\mathrm{SED}}$, and distance $d$ and reddening $E(B-V)$, with their associated uncertainties $\sigma_d$,  $\sigma_{\mathrm{E(B-V)}}$; for this work, we fix $h_0=119$ pc. We give descriptions of each parameter and associated bounds in Table~\ref{tab:modelparameters}.

\subsection{Joint SED and LC model}\label{subsec:joint}

We couple the isochrone interpolator to the light curve solver for simultaneous SED + LC fitting. During a joint fit, the SED model provides binary radii and \kepler bandpass fluxes, which are used as inputs to the LC module; this effectively reduces the number of free parameters from 13+$N_{\mathrm{quarters}}$+10 to 20+$N_{\mathrm{quarters}}$. The full set of
parameters is listed in Table \ref{tab:modelparameters}. We express the joint likelihood function as the product of likelihoods given all sets of data (e.g., LC, SED, extinction, and distance), assuming that each data point is uncorrelated and uncertainties are Gaussian: 

\begin{equation}
\begin{aligned}
\log L = -\frac{1}{2}\sum_i \left(\frac{\Delta \mathrm{LC}_i^2}{(\sigma_{\mathrm{LC}, i}^2 + \sigma_{\mathrm{LC, sys}}^2)} + \log(\sigma_{\mathrm{LC}, i}^2 + \sigma_{\mathrm{LC, sys}}^2)\right) 
- \frac{1}{2}\sum_j \left(\frac{\Delta \mathrm{SED}_j^2}{(\sigma_{\mathrm{SED}, j}^2 + \sigma_{\mathrm{SED, sys}}^2)} + \log(\sigma_{\mathrm{SED}, j}^2 + \sigma_{\mathrm{SED, sys}}^2)\right) 
\\- \frac{1}{2} \left( \left(\frac{\Delta d}{\sigma_d} \right)^2 + \log(\sigma_d^2)\right) - \frac{1}{2} \left( \left(\frac{(\Delta \mathrm{E(B-V)}}{\sigma_{\mathrm{E(B-V)}}} \right)^2 + \log(\sigma_{\mathrm{E(B-V)}}^2)\right) + C\ , 
\end{aligned}
\end{equation}
Here, $\Delta$LC, $\Delta$SED, $\Delta d$, and $\Delta$E(B-V) are the light curve, spectral energy distribution, distance, and reddening fit residuals, respectively. The distance and E(B-V) uncertainties $\sigma_d$ and $\sigma_{\mathrm{E(B-V)}}$ are allowed to float. The systematic error terms for LC and SED are added in quadrature to the observed uncertainties, and importantly the log quantities penalize inflating the error bars. This prevents the model from converging to a poor fit compensated by over-inflating the error bars, and allows the model to weight the LC or SED differently, if the data prefers it. The best-fit solution corresponds to the maximum likelihood model.

\subsection{Model initialization, optimization, and Bayesian parameter estimation}
\label{subsec:fitting}

Due to the non-linear nature of the SED and light curve models, we fit the time-series and multi-band photometry in three stages:
\begin{enumerate}
    \item A fit to only the \kepler light curve.
    \item A fit to only the spectral energy distribution.
    \item A combined fit to the \kepler light curve and spectral energy distribution, using the prior stages to initialize the fit.
\end{enumerate}

For each stage, we optimize the solutions via \texttt{lmfit} \citep{lmfit}, a non-linear least squares minimization algorithm based on the Levenberg-Marquardt algorithm \citep{More1978}. We find that the optimization routine is highly sensitive to initial parameters; for initial solutions far away from local minimum, the optimization scheme may become stuck, or tend toward boundary values. Consequently, we initialize the optimizer with a combination of gridded parameters and parameter guesses from observables. We describe our fitting procedures in practice below.

For LC fitting, we use observed eclipse depths, durations, and times to estimate initial parameters, and pad these guesses with a grid in radius ratio, \ecosw, \esinw, and impact parameter. We then optimize each set of initial parameters and save the solution with the smallest chi-square value. For SED fitting, we use a brute-force grid method to find the best-fit SED model(s) with stellar parameters consistent with the light curve solution: we initialize each set of SED fits in a grid of $M_{\mathrm{sum}} \in \{0.5, 1.0, 1.5, 2.0, 2.5, 4.0\} M_{\odot}$, $Q \in \{0.1, 0.3, 0.5, 0.9\}$, $z$ value from KIC, $E(B-V)$ value from \cite{Schlafly2011} and $d$ from \cite{Bailer-Jones2018}. Then, for each set of $M_1, M_2, z$, we determine the upper age limit given the initial mass in the PARSEC model, and evenly sample age values in log space. For each nested subgrid, we optimize the fit via \texttt{lmfit} and save solutions with the lowest \chisq\ value.

The best-fit parameters from the separate SED and LC solutions seed the joint-fit optimization. When radii values from SED optimization are in tension with LC-derived ones, we pad the separately derived best-fit parameters with a grid of values in mass sum, mass ratio, and age, and step through a grid of these initial parameters for simultaneous SED+LC fitting.

To quantify uncertainties and degeneracies in the model, we use the best-fit solution from the joint SED+LC optimization to seed Monte Carlo Markov Chain (MCMC) simulations. We use an affine-invariant MCMC ensemble sampler, \texttt{emcee} \citep{Foreman-Mackey2013}, with 128 walkers to sample the posterior distribution of parameters for each binary. We place log-uniform priors on the systematic error terms, Gaussian priors on $d$ and E(B-V), and flat priors on all other parameters; see Table~\ref{tab:modelparameters} for details on parameter bounds. In addition, uniform sampling of age and eccentricity, under parameter transformation to $\tau=\log_{10}$(age [yr]) and ($e \sin\omega, \ e \cos\omega$), requires placing a prior of $\exp(\tau\ln10)/\ln10$ and $1/e$, respectively. As the model already contains a large number of free parameters, we fix the crowding parameters during posterior sampling to the values found from the optimization stage. 

We generate Markov chains with 128 walkers for $N_{\mathrm{iter}}$ = 100,000 -- 700,000 iterations, discarding chains before burn-in for posterior analysis. We define the burn-in period as when the chains cross 5 times the median log-likelihood value of the entire ensemble population, adapting the methodology of \cite{Tegmark2004}. Because the chains are initialized from a high-dimensional Gaussian distribution centered on the \texttt{lmfit} optimized solution, burn-in is typically low ($\lesssim10,000$ iterations), and a long burn-in period may signify that the optimized solution was located at a local minimum. For systems exhibiting this behaviour, we restart the Markov chains using the previous run's maximum-likelihood solution.

We use multiple diagnostics to assess convergence. First, we check that the acceptance fraction is between 0.01 and 0.4. Next, we estimate the integrated autocorrelation time, $\tau_{\mathrm{acf}}$, of the ensemble and verify that it is appropriate for the chain length. Finally, we visually inspect that the chains appear well-mixed in trace plots. Longer autocorrelation times typically correlate with smaller acceptance fractions, and large $\tau_{\mathrm{acf}}$ values may indicate a parameter is not well constrained by the data, the parameter space is multi-modal, or that the observed binary cannot be fully captured by co-eval model isochrones. We note that for shallowly eclipsing systems or those exhibiting tidal distortions, the autocorrelation times for $b$, age, metallicity, and mass may be long, with effective chain length $N_{\mathrm{iter}} / \tau_{\mathrm{acf}}$ = 15 -- 20, even after $>$500,000 generations of 128 walkers. This is because grazing geometries do not well constrain the light curve model, and isochrone models do not well predict stars modified by additional physics. \edit{Rather than excluding these EBs from our final sample, for completeness, we report their parameter estimation and discuss at length the factors that limit those results in \S\ref{subsec:mass} and  \S\ref{subsec:limitations}.}  In contrast, flat-bottomed light curves indicative of total or annular eclipses completely break any degeneracy with inclination. For non-grazing, well-detached systems, our median chain lengths are $>$50 times $\tau_{\mathrm{acf}}$. We find that period, times of eclipse, and $e\cos\omega$ converge well for all EBs. 

\subsection{Radial Velocities}
\label{subsec:flavien_rvs}
In order to validate our method, we compare our inferred stellar masses against RV masses from existing literature studies (see \S \ref{sec:discussion}). To supplement the number of RV comparison targets, we mine spectroscopic data from The Apache Point Observatory Galactic Evolution Experiment (APOGEE; \citealp{Majewski2017}), and extract RVs for three EBs in our sample: KICs 5284133, 5460835, and 6610219. They compose a small overlap of double-lined spectroscopic systems (SB2s) with $\ge$6 number of APOGEE visits and were not being studied by the EB Working Group. 

The radial velocities of the SB2 spectra are derived using a multi-order version of the Two-Dimensional Cross-Correlation algorithm called \verb+TODMOR+  \citep{Zucker1994, Zucker2003, Kiefer2018}. We refer the reader to \cite{Kiefer2016} for a full description of the method and summarize the technique in brief here. First, we normalize each APOGEE spectrum by a percentile rank-filtered version of itself \citep{Hodgson1985, Faigler2015, Halbwachs2016}. Empirically, we fix the percentile at the 75\%-level and the window size at 50\,\AA, a compromise to suppress undesirable deformation of the pseudo-continuum but conserving the narrower stellar lines. For each target, the spectra with the largest separation of the components are then matched by $\chi^2$-optimization with 2 synthetic spectra from the PHOENIX library \citep{Hauschildt1999}. The synthetic spectra are also normalized with the same procedure, and broadened to match the line spread function of APOGEE. Averaging over all selected spectra, we obtain stellar parameters for both components of each binary system, including effective temperature, surface gravity, metallicity, flux ratio at 1.6$\mu$m, and $v\sin i$. Finally, we calculate the two-dimensional cross-correlation function of each observed spectrum and the two synthetic spectra, as explained in \cite{Zucker2003}, to derive the RVs of both binary components simultaneously.

\section{Results}
\label{sec:results}

We present the results of our simultaneous SED + LC modeling of 728 \emph{Kepler} EBs. We find that the majority of our models are good fits to the data. In Tables \ref{tab:ebresults1} and \ref{tab:ebresults2}, we list orbital and stellar solutions for a small subset of EBs in our sample, respectively; these inferred binary parameters correspond to the 50th percentile values of the MCMC chains after burn-in, with error bars corresponding to 16th and 84th percentile values. 

The formal uncertainties on mass, age, and metallicity, likely underestimate true, physical uncertainties, as they are reliant on the input stellar isochrone models. \kepler light curves, with their continuous, high-precision 30 min cadence data, offer exquisite constraints on the timing and shapes of eclipses, which map onto the precision of retrieved orbital elements. The inferred geometric elements are particularly precise if the system exhibit flat-bottomed primary and/or secondary eclipse, as total or annular eclipses break degeneracies associated with inclination. For low signal-to-noise or shallow eclipses, however, there is less constraining power in parameters that influence the duration and shapes of the eclipses, such as $e\sin\omega$ vector, impact parameter (inclination), radius ratio, and flux ratio. This degeneracy is a general problem in eclipse modeling, and may be ameliorated by additional spectroscopic (e.g., flux ratios if systems are SB2s) or astrometric constraints. We note that about 10\% of our total EB sample have shallow eclipse depths (PE and SE depths $<$ 0.05), which are susceptible to degeneracies associated with inclination, and thus may not have unique solutions. 

\begin{figure*}
\includegraphics[width=1\textwidth]{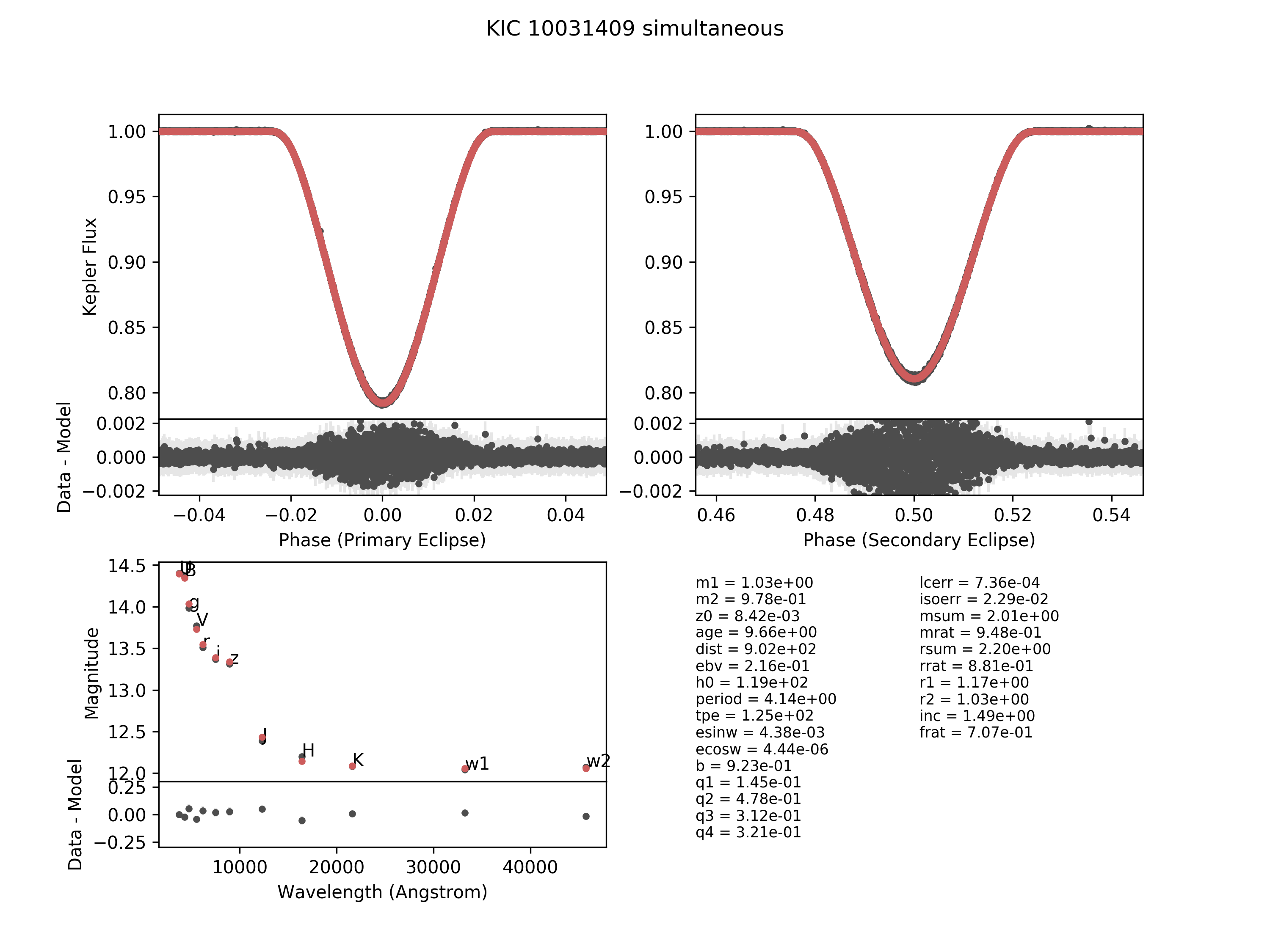}
\caption{Maximum-likelihood (ML) solution for KIC 10031419, a G-dwarf binary in a $\sim$4 d, circular orbit. \edit{The upper panels show the normalized, detrended, and phase-folded primary and secondary eclipse flux data (black) and the light curve fit corresponding to the ML model (red), while the bottom left panel shows the SED fit.} Data points with \kepler\ quality flags $>$8 are masked for visual clarity here (although they are not removed during fitting, see \S \ref{subsec:kepler_data}). The ML parameters are reported for reference. Both the light curve and SED residuals belong to the mean and mode bin from Figure~\ref{fig:eb_medianresiduals}, and indicate a good fit to the data; the scatter during eclipse is consistent with starspot modulations.}
\label{fig:10031409} 
\end{figure*}

\begin{figure*}
\includegraphics[width=1\textwidth]{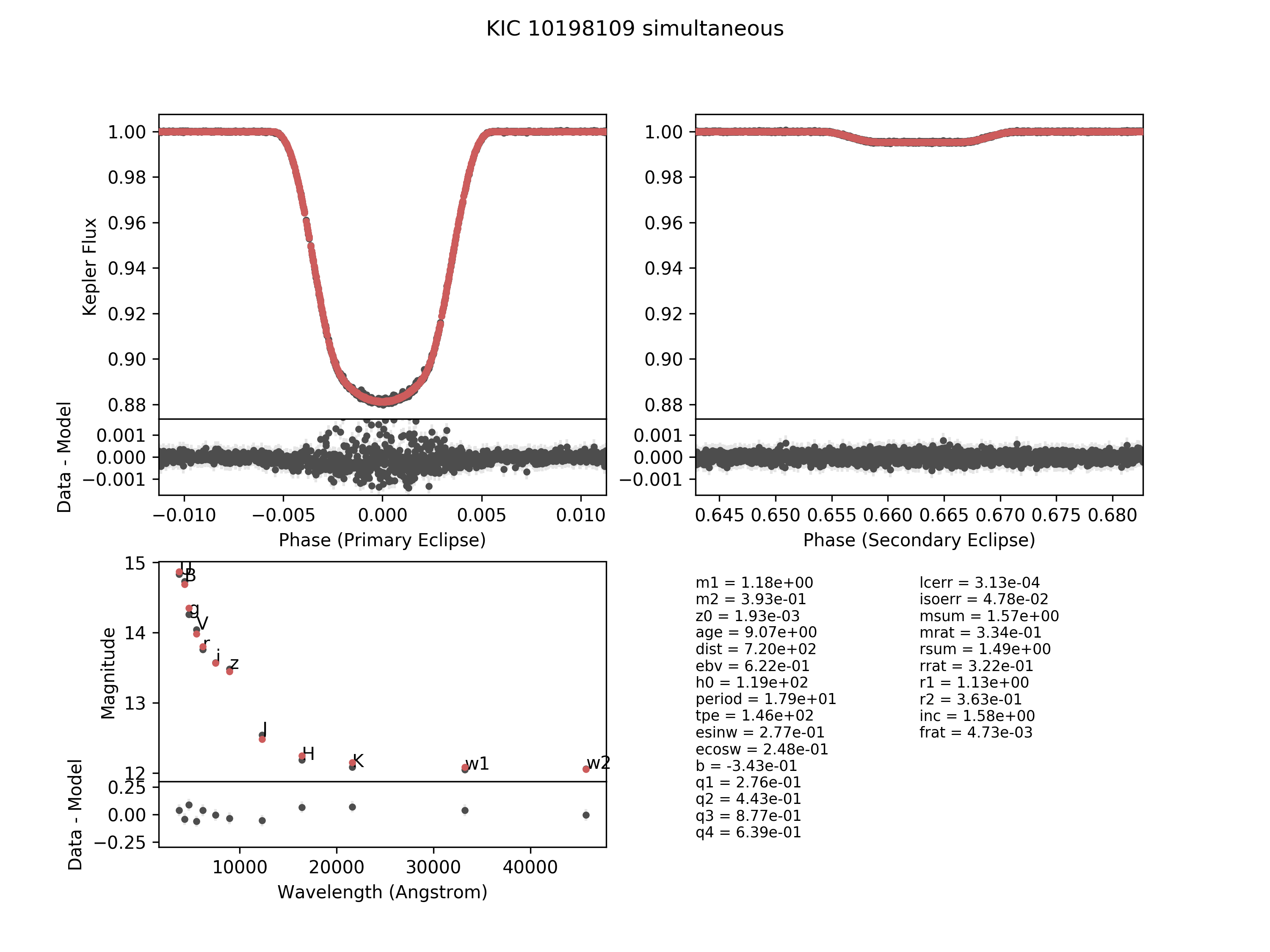}
\caption{Same as Figure \ref{fig:10031409}, except for KIC 10198109, an eccentric, 18 d binary with $Q\sim0.3$. The in-eclipse scatter is small, $\lesssim$1 ppt, and again likely due to starspots.}
\label{fig:10198109} 
\end{figure*}

\begin{figure*}
\includegraphics[width=1\textwidth]{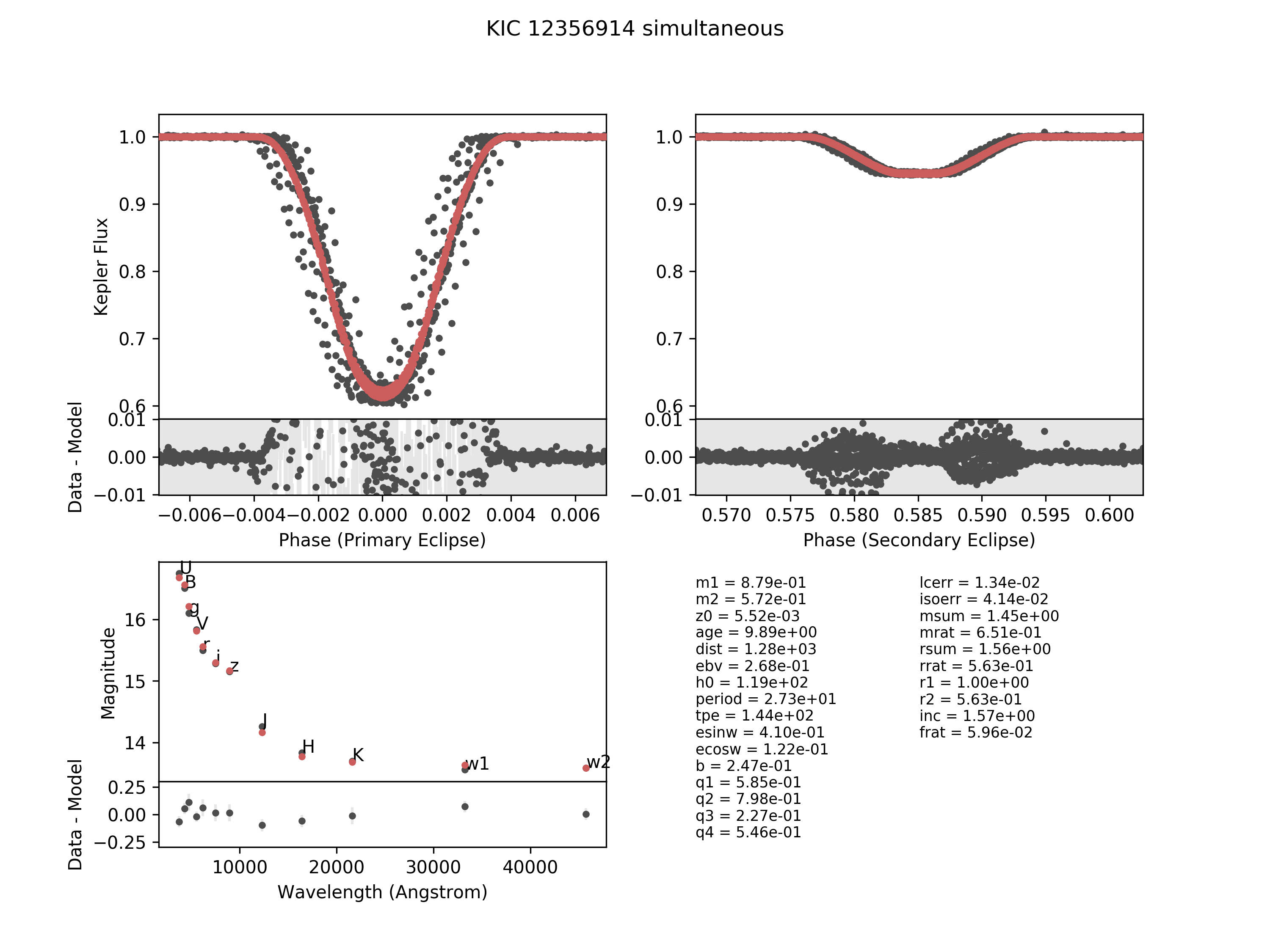}
\caption{Same as Figure \ref{fig:10031409}, but for KIC 12356914, a pair of low mass stars in an eccentric, 27 d orbit around each other. The strong eclipse timing variations present in primary and secondary eclipse pull $\sigma_{\mathrm{sys, LC}}$ toward very large values, and indicate the presence of a third companion in a wider orbit around the central binary, consistent with the more rigorous analysis of \citet{Borkovits2016}.}
\label{fig:12356914} 
\end{figure*}

\begin{figure*}
\includegraphics[width=1\textwidth]{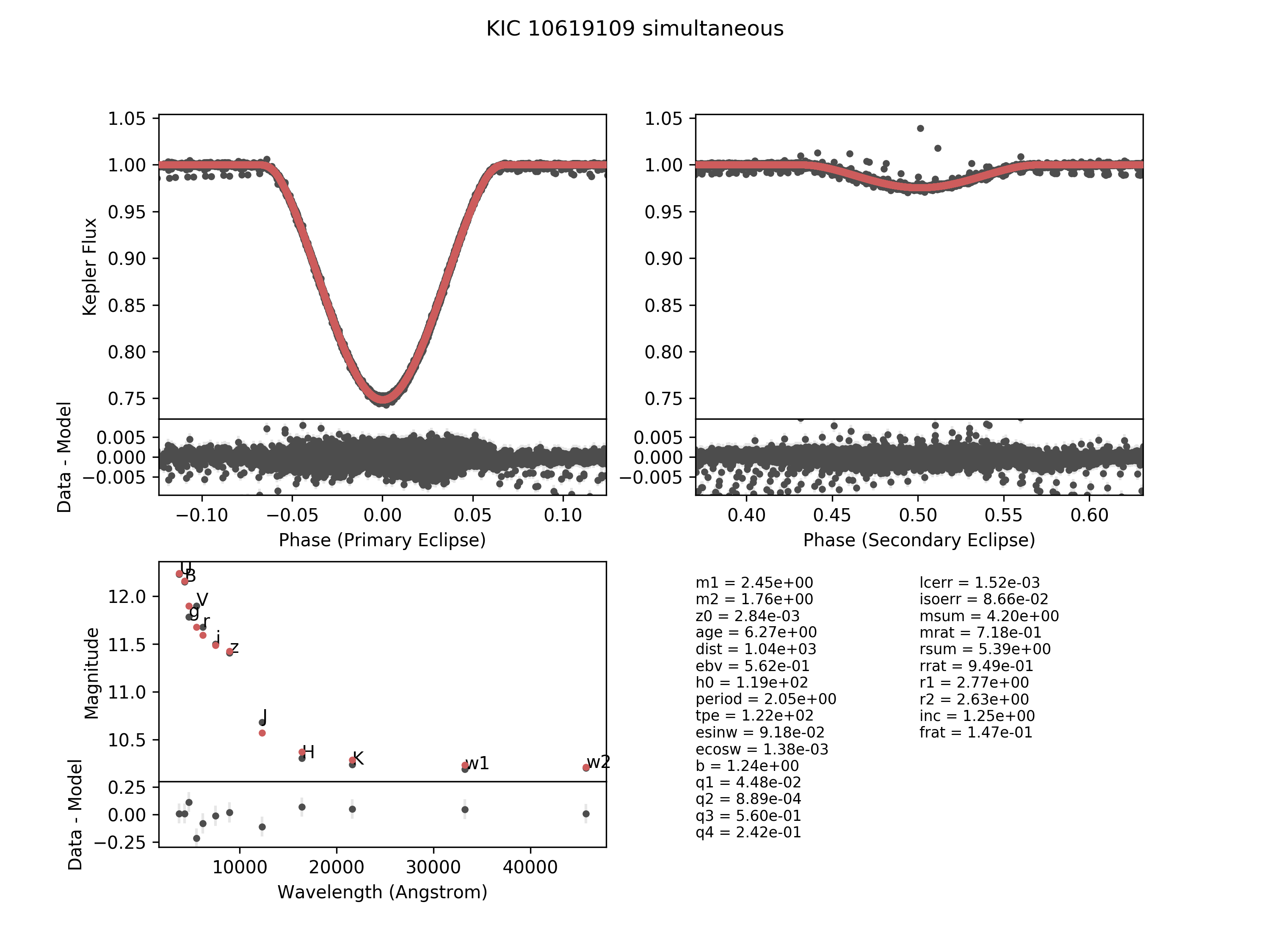}
\caption{Same as Figure \ref{fig:10031409}, but for KIC 10619109, a 2 d, nearly circular orbit binary. Note the relatively large light curve scatter out of eclipse and 0.2 mag scatter in the SED residuals. This is an example of a marginally good fit, but with anomalously young ages, relatively high mass components, and morph parameter $>$0.5, indicative of an Algol-type system (see \S \ref{subsec:age} for further discussion). The outlier data points are an artifact of poor polynomial fitting, due to missing data near/during a particular eclipse. }
\label{fig:10619109} 
\end{figure*}

Furthermore, the 50th percentile values for each parameter frequently correspond to a model which is a poor fit to the data, because the probability distributions are highly asymmetric; therefore, we also report parameter values corresponding to the maximum likelihood (ML) model in Table \ref{tab:ebresults_ml}. We show the ML fits to the SED and light curves, for 5 targets in Figures~\ref{fig:10031409} to \ref{fig:12644769}; these exhibit a range of orbital and stellar parameters, and demonstrate both the versatility and pitfalls of our modeling technique. The full suite of data products for all EBs modeled, including maximum likelihood fits \edit{and posterior parameter distributions plots}, are available in the Github repository.

\subsection{Goodness of fit}
\label{subsec:residuals}
To quantify the goodness of fit for our EB sample using our technique, we measure median absolute scatter between our models and LC+SED data. Figure~\ref{fig:eb_medianresiduals} shows the distribution of residuals for the time series \emph{during eclipse} (top) and multi-band (bottom) photometric fits, in log space. For a majority (75\%) of EBs in our sample, the median residuals are $\lesssim$1 parts per thousand and $\lesssim$0.1 mag for light curve and SED, respectively, indicating that in general model solutions well describe the data (see, e.g., Figures~\ref{fig:10031409}, \ref{fig:10198109}, and \ref{fig:12644769}). We quantify light curve jitter or ``white" noise by computing the median absolute difference (MAD) in out-of-eclipse flux, and overplot the MAD distribution (red dashed line) against the residual distribution for comparison. Note that the MAD noise level computed here typically underestimates true noise in the \kepler light curves, because it does not account for correlated noise due to e.g., spot modulation, Doppler beaming, and third body perturbation (see e.g., Figure~\ref{fig:12356914}). Nevertheless, the majority of systems have model residuals comparable to the white noise, while about 25\% of modeled systems appear to have residuals larger than measured jitter. Systems with morphology parameter $>$0.5 (shown in orange) systematically exhibit larger residuals in the light curve and in the SED; these are typically short period ellipsoidal or interacting binaries (see, e.g., Figure~\ref{fig:10619109}). We further discuss sources of noise that may cause additional scatter during eclipse, as well as sources for the long tail of large residual distributions in \S\ref{subsec:limitations}. 

\subsection{Orbital properties}
The observed sky-projected orbits of binaries can be described by their period $P$, eccentricity $e$, argument of periastron $\omega$, inclination $i$, and semi-major axis $a$. We generate population statistics and histograms using the 50th percentile MCMC posterior values. These distributions provide consistency checks and insight into the physical processes that shaped the binary systems' dynamical states and histories, in the context of observation and selection biases. To fit each distribution, we compute the 50th percentile x-axis values in each bin, and assign Poisson error to bin counts. We adopt this method to generate histograms over convolving posterior probability densities for simplicity, and confirm that this approximation is valid by checking that the 1$\sigma$ uncertainties for each system are much smaller than the histogram bin width. Figure~\ref{fig:EB_histograms} shows the resulting distributions for relative orbital size, eccentricity, inclination, and period. 

The binary period distribution of our sample (upper left panel of Figure~\ref{fig:EB_histograms}) peaks around $P\sim4$ d and follows a gamma distribution with $\alpha=7.6, \beta=4.7$. The shape reflects our target selection process; for reference, the period distribution of the entire \edit{VKEBC} is shown in dark orange and given half weight to keep the y-axis within range. The decline in the number of $P<3$ d EBs in our sample relative to the full catalogue is a direct consequence of selecting for \texttt{morph}$<$0.6 binaries, i.e., at least semi-detached (SD) systems, as over-contact systems are more likely found around binaries with small separations; by contrast, the underlying \kepler EB distribution peaks around $P\sim0.5$ d. 

The inclination distribution is shown in the top right panel, where we fold the values about 90$^{\circ}$. The distribution peaks around $\sim$edge-on, as expected for eclipsing geometries, and can be roughly fit by an exponential distribution with $\lambda=0.28$. The tail of the distribution is somewhat heavy because a large portion of the EBs in our sample have short periods ($P\approx1-2$d), which allows for smaller inclination values (less edge-on) to satisfy the condition for eclipse: $\vert a\cos i \vert \le R_1 + R_2$ for circular orbits. We expect this tail may be even larger if we relaxed the \texttt{morph} $<0.6$ criteria, as higher morphology values correlate strongly with shorter periods and thus higher eclipse probability.

The eccentricity distribution (bottom left panel) cannot be described by a single population, as evidenced by the bimodal distribution in $\log e$ with peaks around $-1.6$ and $-0.5$ ($e \approx 0.03, 0.3$, respectively). The excess of $\sim$circular binaries is expected, given the predominance of shorter period $P<10$ d binaries in the sample, which are typically tidally circularized; this peak near circular orbit would be even stronger if we included the full \edit{VKEBC} sample, which contain more short period EBs. The lower right panel of Figure~\ref{fig:EB_histograms} shows the distribution of binary orbital separations as the sum of stellar radii relative to the semi-major axis, $\frac{R_1 + R_2}{a}$. This is to first order a proxy for light curve morphology; smaller values indicate the stars take up a small fraction of the average orbital separations, while larger values mean the sizes of the stars are comparable to their orbital separation. In the latter case, the stars would experience large tidal forces, resulting in orbital circularization and ellipsoidal deformation of stellar surfaces. Indeed, EBs in our sample with higher morphology values (\texttt{morph}$>0.5$) represent the bulk of systems with large fractional radii. The turnover at larger $\frac{R_1 + R_2}{a}$ values is a consequence of our input criteria to exclude \texttt{morph}$>$0.6 binaries, which are more likely to be tidally interacting.

Figure~\ref{fig:EB_Pcorrelations} illustrates the morphology (top), inclination (middle), and eccentricity (bottom) of EBs in our sample as a function of their period. Morphology parameter values are truncated at our input cut-off of 0.6, and decline with increasing period, such that $P\gtrsim10$ d systems are typically well-detached binaries. This trend encapsulates the effect of tidal forces which scale inversely and steeply with orbital separation. Shorter period binaries ($P<5$ d) have higher ellipsoidal amplitudes and are more likely to interact, and thus have larger morphology values in general. Roughly 18\% of our total sample have \texttt{morph}$\ge$0.5. 

As expected, binary inclination tapers to $\sim$edge-on (90$^{\circ}$) at longer orbital periods (middle panel), due to observational bias in the geometry of detecting eclipses. The orbital separations of very short period binaries ($P<4$ d) are small enough that grazing eclipses still occur for orbits that are inclined significantly with respect to the observer ($i\sim70^{\circ}$). Eclipsing binaries that harbour circumbinary planets \citep{Doyle2011, Welsh2012, Orosz2012a, Orosz2012b, Schwamb2013,Kostov2014,Welsh2014,Kostov2016} are highlighted as orange stars; they are well-aligned systems with $P\approx10-50$ d. 

The eccentricity-period plot (bottom panel) shows the prevalence of tidal circularization in binaries with orbital periods $\lesssim$10 d. This result is consistent with findings from previous studies of \kepler binaries \citep{VanEylen2016, Price-Whelan2018}, although we do not recover a significant difference in the fraction of eccentric binaries with respect to hot and cold stellar components \citep{VanEylen2016}. 

\subsection{Stellar properties}
\label{subsec:stellar}
Next, we present the distributions of stellar parameters ($\tau = \log_{10}$(age [yr]), mass ratio $Q$, surface gravity $\log g$, and temperature $T$) of our EB sample in Figure~\ref{fig:EB_histograms_stellar}. The age distribution (upper left panel) is bimodal, with a strong peak between $\sim$1-10 Gyr as expected, and another broad but much smaller distribution at younger ages, centered around $\tau\sim$7, or 10 Myr. Were the stellar distribution drawn from a uniform distribution in formation timescale over 10 Gyr, which approximates the star formation of the Galactic disk from which this sample is drawn, then we would only expect $\approx$1\% of stars with ages $<100$ Myr and $\approx$0.1\% with ages less than $10^7$ yr, rather than the $\approx$10\% for $\tau<7$ and $\approx$25\% for $\tau<8$ which comprise our sample. The over-representation of young ages corresponds to instances where single-star isochrones poorly predict the observed stellar properties, and we discuss the age results further in \S \ref{subsec:age}

The $\log g$ distribution (upper right panel) shows that the majority of detached EBs in the \kepler sample are dwarfs ($\log g\sim$4.5), although about 10\% of them have started to evolve off the main sequence as subgiants with $\log g < 4$. We decompose the $\log g$ distributions of primary (dark grey) and secondary (light grey) components, and stack them vertically. The $\log g$ distributions for EB primaries and secondaries peak at slightly different values: 4.2 and 4.5, respectively. For reference, we overplot (in dark orange) the distribution of $\log g$ from the full \kepler Objects of Interest catalogue (KOI), i.e., single stars. The combined (primary + secondary) $\log g$ distribution is similar to the underlying KOI $\log g$ distribution in that both KOI and KEBLAT EB surface gravity distributions $\log g \sim 4.5$, consistent with main sequence dwarfs. However, there is a discrepancy between KOI and overall KEBLAT EB samples in the $\log g \approx 3.8-4.2$ range, where slightly evolved sub-giants are overrepresented in our sample compared to KOIs. The $\log g$ distribution for our sample's primary stars also peaks at a lower $\log g$ value than the full KOI sample. These differences may not be significant, given typical uncertainties in KOI $\log g$ values are 0.4 \citep{Brown2011}, although it may be related to the higher eclipse probability for stars of larger radius. 

The mass ratio distribution ($Q=M_2/M_1$; lower left panel) shows a general positive trend. Typically, $Q>1$ systems indicate one or both of the components are not normal stars, i.e., non-MS or mass exchange has modified a system's nascent binary mass ratio. For plotting purposes here, however, we invert the mass ratios for $Q>1$, and stack the resulting histogram on top of the $Q\le1$ population, so that the values are in the range [0, 1]. The combined distribution contains a dearth of low-mass companions ($Q<0.3$) and strongly favours similar-mass binaries ($Q\sim0.95$), consistent with spectroscopic results of solar-type binaries by \citet{Raghavan2010} and \citet{Tokovinin2006}. \edit{This increasing $Q$ trend, \emph{if physical}, should reflect formation values, as our morphology cut disfavours semi-detached binaries and excludes overcontact binaries, e.g., W Uma systems, which are able to exchange mass and evolve away from their nascent $Q$ values (see, e.g.,  \citealp{Rucinski2001,Yakut2005,Gazeas2006}). We further discuss the reliability of mass ratio inference in \S\ref{subsec:mass}.}

\edit{ Here, we briefly consider how observational biases (e.g., eclipse probability) and the input selection criteria (e.g., \texttt{morph}$<$0.6, SE depths) may affect this apparent mass ratio trend; quantifying the combined selection function of \emph{Kepler}, the VKEBC, and our sample is beyond the scope of this paper. Since eclipse probability and morphology values decrease with increasing period (see Figure~\ref{fig:EB_Pcorrelations}), we examine the distribution of mass ratios in different period bins. } The lower right panel of Figure~\ref{fig:EB_histograms_stellar} shows $Q$ distributions (for $M_2/M_1\le1$ only) for very short ($P<4$ d; blue solid), short ($4\le P <10$ d; orange dashed), and long ($P>10$ d; green dotted) period binaries, where mass bins are widened relative to the cumulative $Q$ distribution to account for low bin counting. Within Poisson counting noise, we find no significant deviations in $Q$ distributions across different binary period bins. \edit{We expect eclipse probability to be relatively uniform across $Q$ and not affect their distribution per orbital period bin, except at very low mass ratios which preferentially produce extremely shallow secondary eclipses. }

\subsection{H-R diagram for \kepler EBs}
\label{subsec:hr}
Using predictions from stellar isochrones based on best-fit SED+LC models, we plot luminosity in absolute \kepler magnitudes, temperature, and metallicity for our EB sample in Figure~\ref{fig:EB_hrdiagram}. The resulting H-R diagram shows a spread in the main sequence, with a smattering of evolved stars, but no prominent red giant branch. The lack of giants is likely due to a combination of bias in isochrone fitting and target selection.  The target selection is affected both by 1) colour cuts in the \kepler catalogue and 2) the morphology cut we made for our sample of $\sim$detached EBs. In the first case, \kepler was designed to detect Earth-size planets around Sun-like stars and selected for FGK dwarfs. In the second case, our morphology cut may bias the sample because red giants are low density, and thus much more easily distorted by gravity and produce (over-)contact morphologies, which are preferentially excluded.  Although there is some giant contamination, the majority of \kepler input catalogue targets are dwarfs. \cite{Berger2018} recently measured precise radii for $\sim$200,000 \kepler stars  using \emph{Gaia} DR2 parallaxes and properties from the DR25 Kepler Stellar Properties Catalog \citep{Mathur2017}, and found that only 23\% and 12\% of the stars are subgiants and giants, respectively. We find $\sim18$\% and $\sim10$\% of the primaries and secondaries in our sample have $\log g<4$ (see upper right panel of Figure~\ref{fig:EB_histograms_stellar}). We discuss how isochrone fitting may bias the age and $\log g$ results for KEBLAT EBs in Section~\ref{sec:discussion}. 

\subsection{ETV systems and triple candidates}
\label{subsec:triples}
\begin{figure*}
\includegraphics[width=1\textwidth]{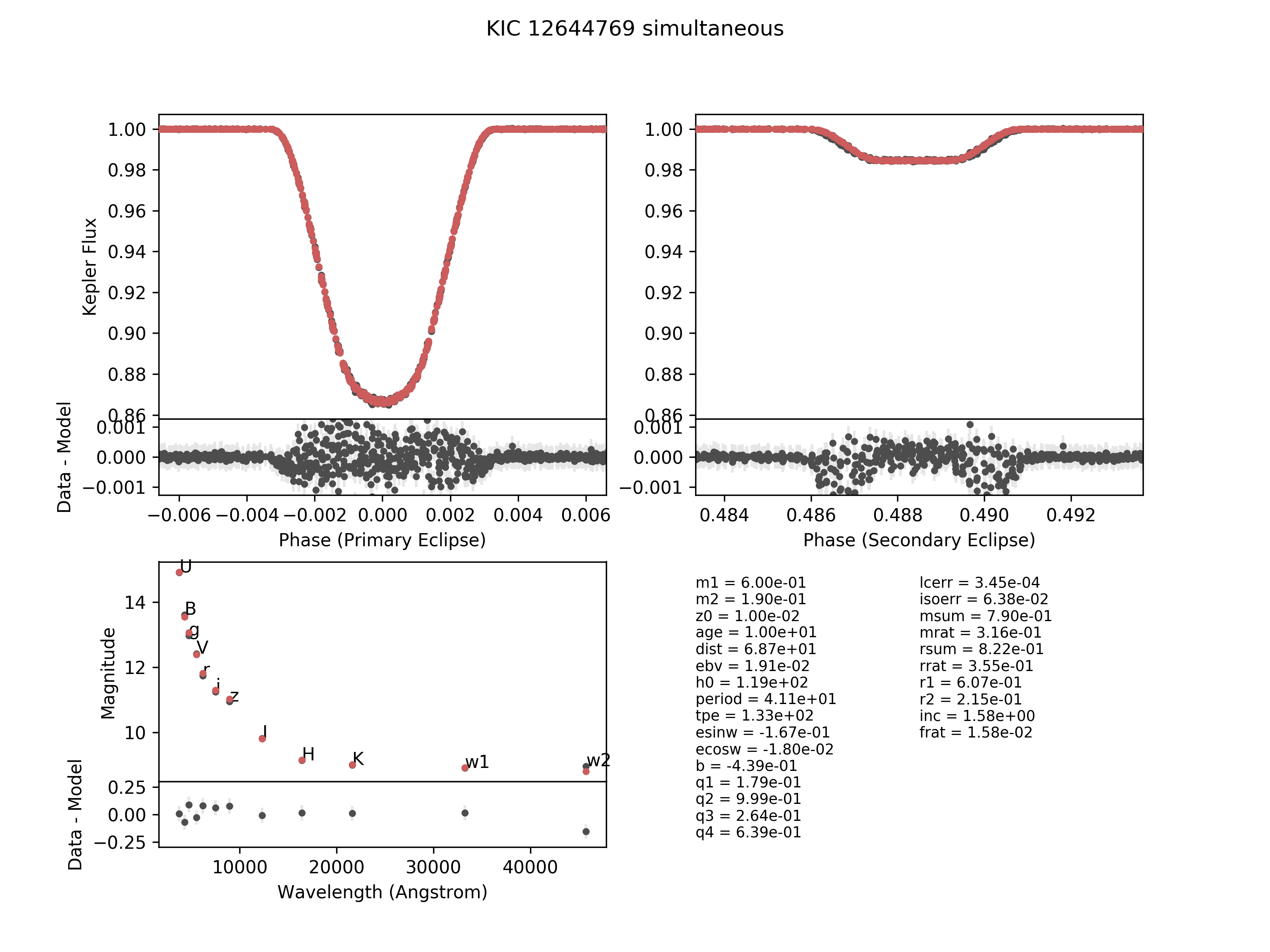}
\caption{Same as Figure \ref{fig:10031409}, but for KIC 12644769, a system in which two M dwarfs orbit each other every 41 days in a slightly eccentric fashion. The scatter in-eclipse is relatively small at $<$1 ppt, but nevertheless exhibit interesting phenomena. Residuals during primary eclipse are consistent with starspot modulations (on the primary star), while residuals at secondary ingress and egress indicate small ETVs by a small tertiary component. Indeed, KIC 12644769, aka Kepler-16, is a known circumbinary planet host \citep{Doyle2011}. }
\label{fig:12644769} 
\end{figure*}

In addition to extracting binary properties, we visually inspected the \edit{light curve fits} of all EBs modeled and identified 84 systems with eclipse timing variations (ETVs), potentially due to perturbations by a tertiary companion or apsidal motion from tidal deformation of binary components. Out of these 84 systems, one is a confirmed circumbinary planet host (i.e.\ Kepler-16/KIC 12644769, see Figure~\ref{fig:12644769}) and 48 have been previously identified and characterized \citep{Gies2012,Rappaport2013, Conroy2014, Orosz2015, Borkovits2016}, leaving 35 newly identified candidates. We list all ETV systems identified in this work in Table~\ref{tab:etv_candidates}, where additional lettering denotes provenance of previous identification, and reported periods, mass ratios, and eccentricities correspond to maximum-likelihood solutions.

The ETVs typically manifest in the light curve residuals as additional scatter around eclipse ingress and egress, with a ``pinching" at the center of eclipse, see, e.g., Figures~\ref{fig:12356914} and \ref{fig:12644769}. Because ETVs cause eclipses to shift back and forth in time, the residuals take on a shape proportional to the derivative of the light curve. They are largest when the slope of the eclipse is steepest but decline to zero mid-eclipse where the slope is zero, which is responsible for the pinch. Stellar pulsations or starspot modulations may confound ETV signals or lead to false positives; however, these stellar variations tend to affect the entire range of the eclipse and take on a shape similar proportional to the shape of the eclipse. We note that because the systems reported here are inspected by eye, they are typically cases with strong ETV signals and do not comprise a homogeneous sample. Both \cite{Conroy2014} and \cite{Orosz2015} refer to a large, comprehensive study of triples in well-detached \kepler EBs still in development by Orosz, in prep, which would be a source of detailed characterization not carried out here.

\subsection{APOGEE-derived Mass Ratios}
As a part of our investigation into the reliability of KEBLAT (non-RV) mass ratio estimates as compared to RV-derived $Q$, we measured and modeled RVs from APOGEE spectra for three SB2 systems in our sample, KICs 5284133, 5460835, and 6610219 (see \S\ref{subsec:flavien_rvs} for a description of the RV extraction and model). This is an independent analysis from KEBLAT, without LC or SED information or priors, and relies on software from \cite{Kiefer2018}. Absolute masses of the individual components require the inclination constraint from the light curve and cannot be derived from the RV alone; however, the mass ratios depend only on the ratio of RV semi-amplitudes. We present mass ratio solutions for the three SB2s in Table~\ref{tab:flavien_rvs}, and show the best-fit radial velocity solutions in Figure~\ref{fig:flavien_rv_solutions}. We discuss these results in the context our larger SED+LC modeling effort in \S\ref{subsec:mass}. 

\section{Discussion}
\label{sec:discussion}
Here, we examine the robustness and limitations of our binary model and results. In particular, we determine the accuracy our EB properties by comparing them to various previous studies published in literature, including RV studies which yield direct measurements on mass (see \S\ref{subsec:reliability}). We are able to reliably infer masses using only photometry for detached, main-sequence binaries. Our solutions are not broadly applicable for EBs with components modified by mass transfer or with red giant companions, as they are not well captured by theoretical isochrones. In \S\ref{subsec:limitations}, we discuss these caveats and devise diagnostics to identify them using observables and model parameters. 

\subsection{Reliability of parameter estimation}
\label{subsec:reliability}
\subsubsection{Mass Measurements}
\label{subsec:mass}

We compile from several radial velocity studies a list of \kepler EBs with published mass solutions common to our sample in Table~\ref{tab:ebcompare}. These masses represent RV ``truths" with which we compare our photometry-only mass inferences. We adopt the mass ratio $Q$, a quantity independent of inclination, to compare our results; this allows us to use the APOGEE-derived mass ratios from our independent RV analysis. 

Figure~\ref{fig:ebQcomparison} compares KEBLAT mass ratios to RV solutions from the literature for an overlap sample of \edit{55} EBs. The majority of data points lie close the one-to-one line, with a cluster of discrepant values around KEBLAT $Q\sim$0.75. We hypothesize that differences in the sub-population of binaries within the overlap sample may explain the varying levels of $Q$ agreement. We use a mixture model to test our hypothesis and quantify the degree of fitness. Following the convention of \cite{Hogg2010}, we include a linear trend (the ``foreground" model) and assume the outliers are drawn from a Gaussian distribution (the ``background" model). We sample the mixture model with \texttt{emcee} and compute the marginalized posterior probability that each data point belongs to the foreground model $P_{FG}$. We use a threshold $P_{FG}=10^{-16}$ to split the sample into ``foreground" (\edit{42} EBs) and ``background" (13 EBs) populations for additional analysis.

We investigate the EBs from the ``background" or outlier distribution and find they cluster toward two distinct groups: (i) binaries with morphology values $>0.5$ (e.g., ellipsoidal or Algol-type binaries) and (ii) binaries with one or more red giant components. These systems are not robustly fit by KEBLAT because of the inherent model assumptions; our method does not treat non-Keplerian effects such as tidal deformations that may be present in higher morphology systems, and it is reliant on theoretical isochrones, which cannot self-consistently describe systems that have undergone mass transfer. Moreover, the post-MS evolutionary phases are short lived, where small perturbations in mass or age can lead to large differences, or discontinuities in radius, making it difficult to capture via isochrone fitting. While KEBLAT successfully retrieved the mass ratios of \emph{some} \texttt{morph}$>$0.5 and red giant binaries, it typically failed to reproduce physically self-consistent solutions. Instead, the solutions tended toward young, $Q>1$  massive stars with large radius ratios to satisfy the light curve eclipse constraints. One underlying symptom of these mass ratio outliers is that they typically have artificially young derived ages; we discuss this in further detail in \S\ref{subsec:age}. 

We perform linear regression on the ``foreground" EB population (e.g., with outliers removed) in our mixture model and use a systematic uncertainty term $\sigma_{Q, \mathrm{sys}}$ to quantify the accuracy of KEBLAT-derived masses with respect to RV solutions. We find that they correlate linearly with slope $m=0.9\edit{5}\pm0.0\edit{6}$, intercept $b=0.0\edit{1}\pm0.0\edit{5}$, and systematic error $\sigma_{Q, \mathrm{sys}}=0.1\pm0.02$. This agreement indicates that KEBLAT mass ratios are reliable to $\sim$0.1 for well-detached, main-sequence binaries. Additionally, we demonstrate the agreement for absolute masses between KEBLAT and RV analyses. Figure~\ref{fig:eb_absolute_mass} shows primary (filled circle) and secondary (open circle) masses in the overlap sample, after removing EBs with red giant components and \texttt{morph}$>$0.5. The inferred masses from photometry-only are in accord with RV values to within 15\% of the mass of the star (represented by the grey region). We note that not all of the RV studies included simultaneous light curve modeling in their analysis, which may introduce $\sin i$ corrections in the RV absolute masses at the 2\% level; this bias is small relative to the scatter between RV and KEBLAT values, and so we do not account for it.  

In addition to the RV studies, a small number of \kepler EBs in our sample overlapped with systems observed by the Trans-atlantic Exoplanet Survey (TrES) and modeled by \cite{Devor2008}. Their method (MECI) was similar to ours, and used ground based light curves and archival 2MASS photometry in conjunction with isochrones to predict binary masses, but does not use parallax as a constraint upon the model. We applied a quality cut to their catalogue, and selected only for overlap binaries with reduced $\chi^2<2$. Figure~\ref{fig:Q_meci} shows the agreement in primary and secondary mass between KEBLAT and MECI analysis; in general, the masses agree to 20\%, despite both analyses relying only on photometry and adopting different isochrones, data sets, and model details.

For mass inference, the comparison results above indicate that masses derived from photometry here are reliable for well-detached, main-sequence binaries, which comprise the majority of our sample. We recommend approaching systems that have morphology parameter $>0.5$, young age estimates ($\tau\lesssim7.5$), and large radius ratio ($R_2/R_1\gtrsim2$; indicative of red giant secondaries) with caution.  

\subsubsection{Orbital elements and stellar radii \& temperatures}
\label{subsec:orbital}
We compare our inferred eccentricities, inclinations, radii, and temperatures to solutions from the RV studies mentioned above. We supplement this comparison analysis with values from coarse ensemble parameter studies using light curve \citep{Kjurkchieva2017} and SED \citep{Armstrong2014} information. Because the literature studies used here have varying degrees of robustness based on their model assumptions and data quality, we only consider their qualitative agreement. In general, KEBLAT solutions agree with published values from small-sample RV studies to within the measurement uncertainties, and share bulk trends with existing larger-scale light curve or SED studies of \kepler EBs. 

Figure~\ref{fig:eb_compare_e_r} shows that eccentricity, inclination, and radii values between KEBLAT and literature studies generally lie close to 1:1 relationships, denoted by the dashed black lines. \edit{The inclination values show good agreement with RV analyses (see lower left panel), with the exception of \cite{Matson2017} values; we suspect this is because their inclinations are based on machine learning results from phenomenological LC modeling rather than a physical LC model. The eccentricity values (upper left panel) are also generally in very good agreement with published RV work. Our values are broadly consistent with those from \cite{Kjurkchieva2017}, although there is large scatter as their approach suffers from using binned phase data and only fitting $e$ and $\omega$ while holding other parameters fixed at approximate relation values (see their \S2). That is, they assume the eccentricity vectors do not correlate with radius, temperature, and other orbital parameters. However, binning or down-sampling the LC can change the eclipse profiles, especially at ingress and egress where the shapes are sensitive to eccentricity vectors, inclination, and stellar radii. Therefore, we expect our full forward modeling on the entire light curve to yield more robust results.}

The upper and lower right panels show that our inferred radii demonstrate very good fidelity to results from RV studies. In particular, the red giant systems \citep{Gaulme2016} show much better agreement in radii than mass (see Figure~\ref{fig:ebQcomparison}); while the eclipse shape tightly constrains binary radii, the isochrone fitting prefers a more massive MS secondary to a low-mass post-MS component to reproduce its large secondary radius. This may be due to large systematic errors in the isochrones for post-MS evolutionary phases \citep{delBurgo2018} and/or insufficient parameter sampling of complex isochrone morphologies (see \S\ref{subsec:age} for further discussion). 

Figure~\ref{fig:EB_Armstrong} compares general \kepler EB properties derived in this study to those reported in \cite{Armstrong2014}. The goal of their work was to generate binary temperatures for the entire \edit{VKEBC}, using Castelli-Kurucz \citep{Castelli2003} model atmospheres to fit SEDs constructed from UBVJHK magnitudes. Their analysis differs significantly from ours; in particular, they did not utilize the light curves, except to use eclipse depths reported in the \edit{VKEBC} to constrain $T_2/T_1$. As shown in the upper left panel, the effective temperatures of primary components inferred by KEBLAT generally agrees with \cite{Armstrong2014} values as well, although there is a discrepant clump near KEBLAT $\log T_1 \sim 4.1$ vs. Armstrong $\log T_1 \sim 3.8$. This latter clump represents the eclipsing Algol population in our sample, previously mentioned in \S\ref{subsec:mass} and discussed further in \S\ref{subsec:age}. \cite{Armstrong2014} used Gaussian priors centered around KIC values, with an adaptive upper limit of 13000K (see their Table 1), or $\log T \sim $4.1, which explains the dearth of $\log T_1>4$. The temperature ratios generally agree (lower left panel), although there is a large scatter. This is to be expected, since \cite{Armstrong2014} inferred the ratio of temperatures based on relative depth of eclipses as reported by the \edit{VKEBC}, whereas here, we derive $T_2/T_1$ from a full forward model based on light curve and SED data. The upper right panel shows that primary radii values, normalized by estimated EB distance, agree well, while radius ratio as derived from KEBLAT and \cite{Armstrong2014} has significant scatter (see lower right panel). In particular, the inset overlay shows that for a small range of KEBLAT $R_2/R_1$ values, there is a large, nearly uniform spread (0.5-1.0) in Armstrong radius ratios, which the authors state are poorly constrained in their study (see, e.g., Figures 4 and 5 and section 5.2.2 in \cite{Armstrong2014}). Furthermore, we note that the data points with very large radius ratio values ($R_2/R_1>2$) are systems with confirmed red giant components \citep{Gaulme2016}.

\subsection{Limitations \& Diagnostics}
\label{subsec:limitations}
We have demonstrated that our models produce good fits to the data for the majority of our sample and accurate parameter inferences for a subset of EBs with RV overlap. Here, we examine our model assumptions, the inherent limitations to our method, and their effects on our results. We recommend using age ($\tau\lesssim7.5$) and morphology (\texttt{morph}$>$0.5) as diagnostics for systems which may be poorly described by our model, and comment on future improvements. 

\subsubsection{Correlated noise in the light curve}
\label{subsec:noise}
\begin{figure*}
\includegraphics[width=0.9\textwidth]{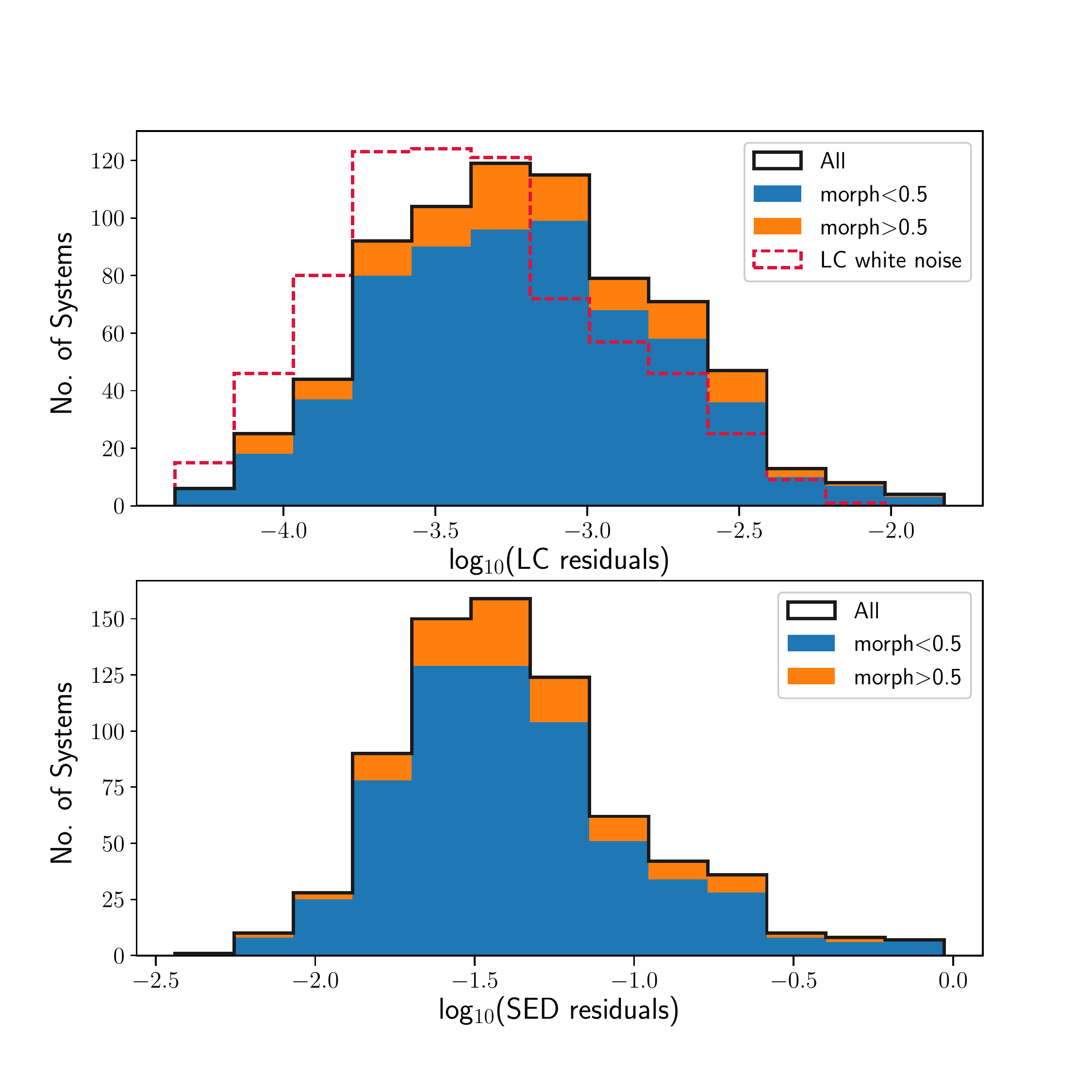}
\caption[width=0.9\textwidth]{The median absolute data-model residual distribution (black curve) for our EB fits in log space; we decompose the distributions between \texttt{morph}$<$0.5 (blue) and \texttt{morph}$>$0.5 (orange) systems and stack them vertically. The majority of systems modeled exhibit residual levels comparable to light curve ``jitter," quantified by the median absolute difference (MAD) in out-of-eclipse flux (red dashed curve). However, this proxy for LC noise does not capture correlated noise well, so light curves which exhibit e.g., large Doppler or ellipsoidal amplitudes, starspot variation, quasi-periodic stellar variability, or third light dilution that varies within each quarter, will have underestimated noise values. This contributes to the difference between in-eclipse model residuals and inherent light curve jitter distributions; systems with large eclipse timing variations, starspot or ellipsoidal variations compose the tail of large residuals ($\gtrsim0.01$). Indeed, \texttt{morph}$>$0.5, e.g., short period EBs likely to exhibit ellipsoidal variations and/or have undergone interactions, have larger residuals relative to \texttt{morph}$<$0.5 systems.} 
\label{fig:eb_medianresiduals} 
\end{figure*}

\begin{figure*}
\includegraphics[width=\textwidth]{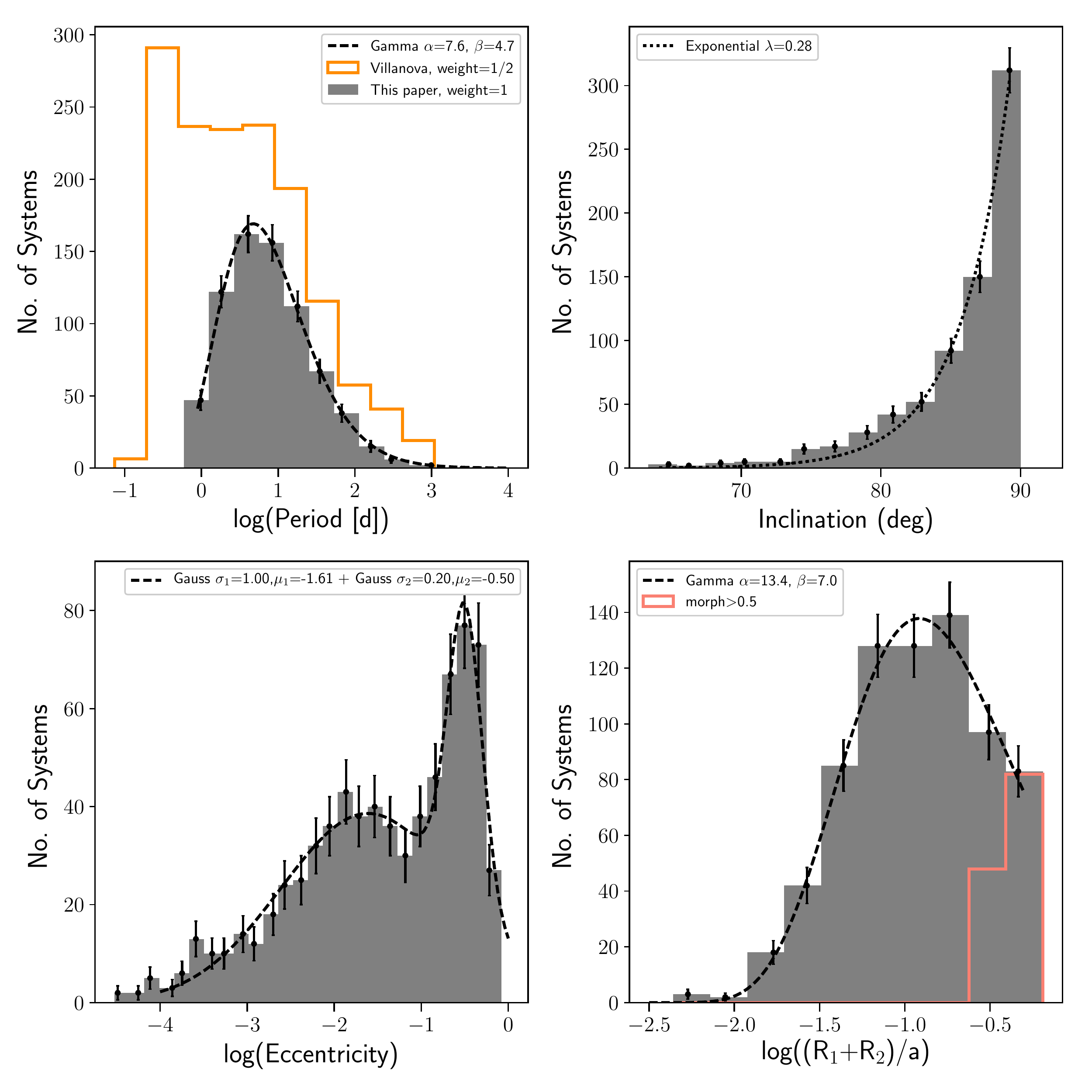}
\caption[width=0.5\textwidth]{Distribution of orbital parameters: period, inclination, eccentricity, and orbital separation relative to stellar size, for EBs in our sample. The period distribution peaks around 3 days, and follows a gamma distribution in shape. The inclination distribution peaks near edge-on and declines exponentially. Note that our fits allow inclinations from 0$^{\circ}$ to 180$^{\circ}$, but here we wish to illustrate the distribution of edge-on vs. grazing systems, so we folded the data about 90 degrees. The eccentricity distribution is bimodal, peaking around $\log e \sim -1.6, -0.5$ corresponding to $e \approx$ 0.03, 0.3. We also plot the distribution of orbital semi-major axis relative to stellar radius, which to first order is a proxy of light curve morphology; about 3/4 of our sample are have $(R_1+R_2)/a<0.2$.}
\label{fig:EB_histograms} 
\end{figure*}

\begin{figure*}
\includegraphics[width=0.8\textwidth]{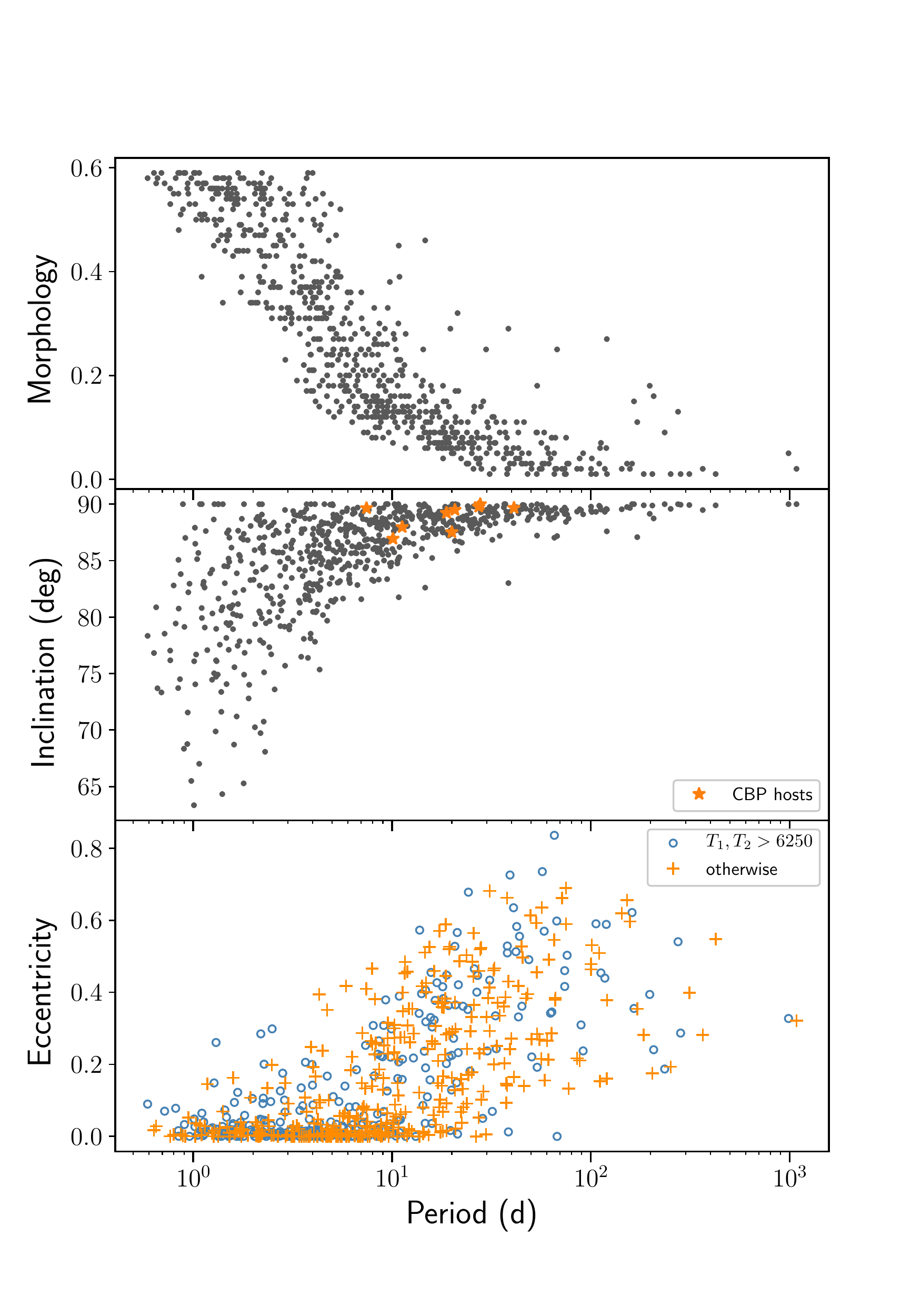}
\caption[width=0.5\textwidth]{Morphology (top), inclination (middle), and  eccentricity (bottom) of EBs in our sample as a function of their orbital period.}
\label{fig:EB_Pcorrelations} 
\end{figure*}

\begin{figure*}
\includegraphics[width=\textwidth]{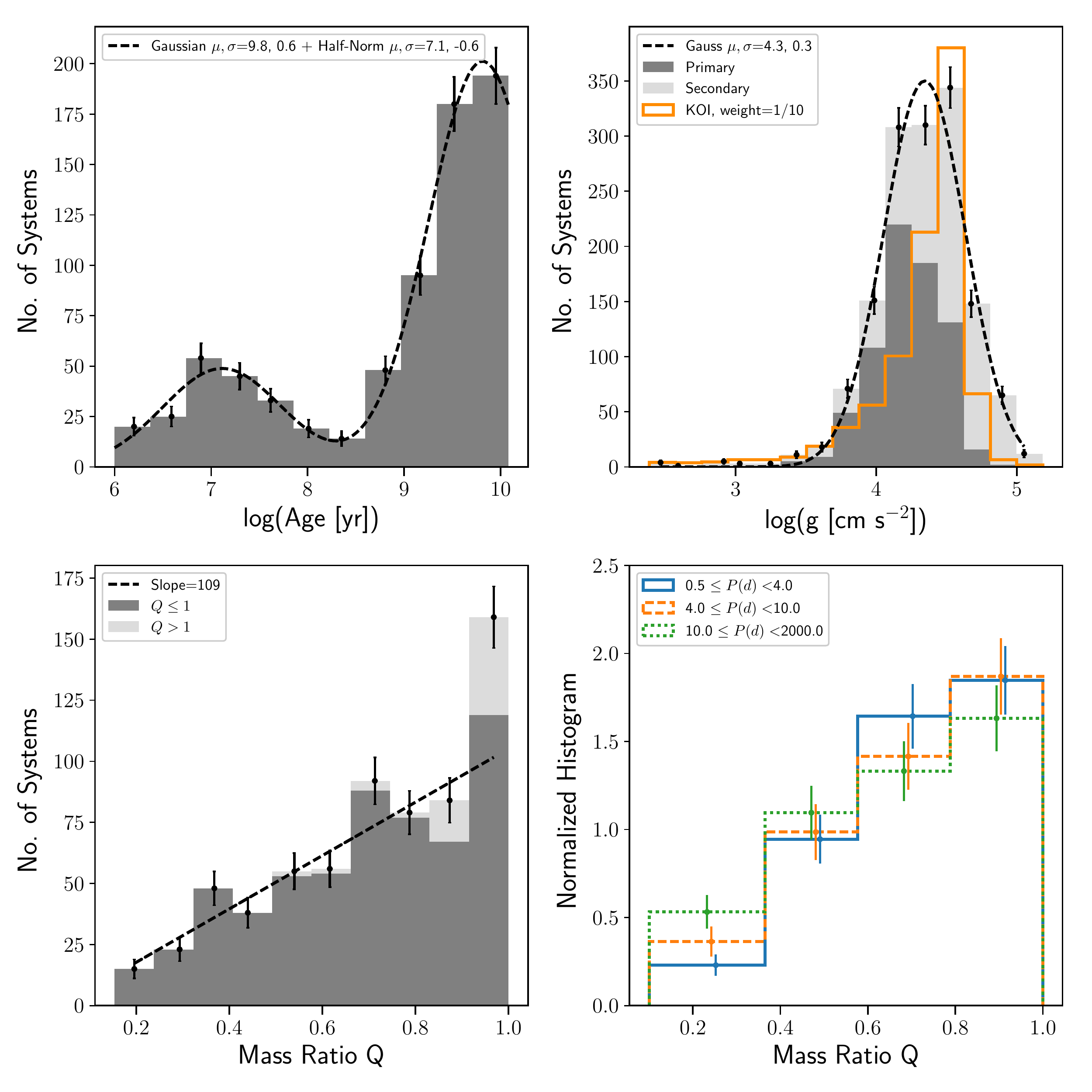}
\caption[width=0.5\textwidth]{Distribution of stellar parameters: age, $\log g$, and mass ratio $Q$, for EBs in our sample. The age distribution (upper left) is bimodal, such that the majority of EBs contain $\sim$billion-year-old stars, with a small but notable excess of young ($<$100 Myr) stars (see text for discussion). The surface gravity distribution (upper right) peaks around $\log g \sim 4.3$, indicating the prevalence of main sequence binaries in our sample. The mass ratio distribution (lower left) shows an increasing slope toward similar-mass binaries, consistent with observations of solar-type binaries in the field \citep{Raghavan2010}. Note that for systems where $Q>1$ (light grey), we invert the mass ratio such that $Q = M_1/M_2$ to keep values in range [0, 1]. The mass ratio distribution for $Q\le1$ (lower right) is relatively uniform across $P<4$, $P=4-10$, and $P>10$ d binaries.


}
\label{fig:EB_histograms_stellar} 
\end{figure*}

\begin{figure*}
\includegraphics[width=\textwidth]{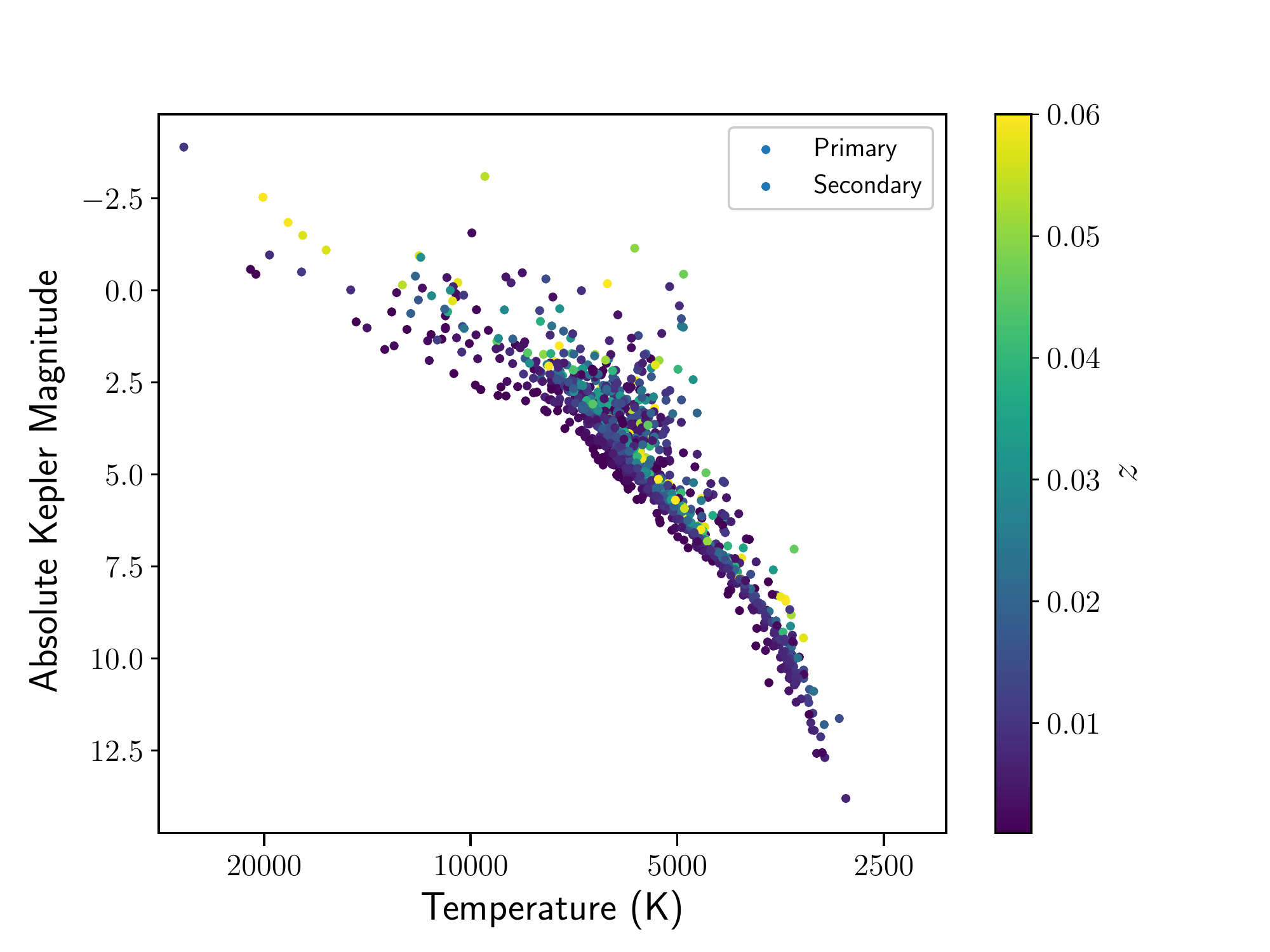}
\caption[width=0.5\textwidth]{EBs from our sample, with temperature and absolute \kepler magnitude of each stellar component, plotted on the HR diagram; the color bar denotes inferred metallicity $z$ of the system. While there is a small population of sub-giants starting to turn off the main sequence, there is an absence of red giant branch. The lack of giants may be due to a combination of isochrone fitting bias and Kepler target selection (see text for discussion).} 
\label{fig:EB_hrdiagram} 
\end{figure*}

\begin{figure*}
\includegraphics[width=\textwidth]{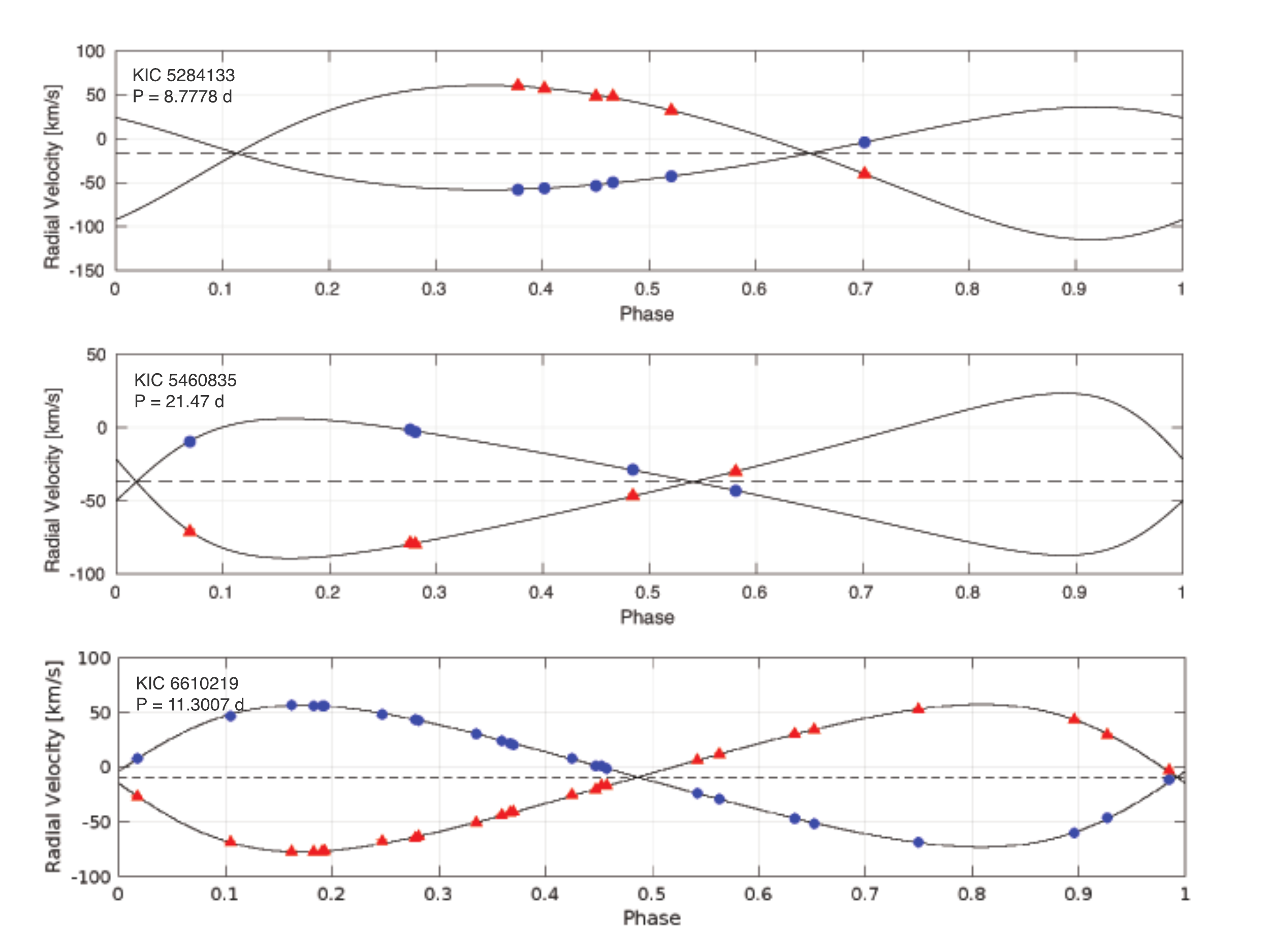}
\caption[width=0.5\textwidth]{Radial velocity solutions in an independent analysis to three SB2 systems in our sample, using the method of \cite{Kiefer2018}.}
\label{fig:flavien_rv_solutions} 
\end{figure*}

\begin{figure*}
\includegraphics[width=0.9\textwidth]{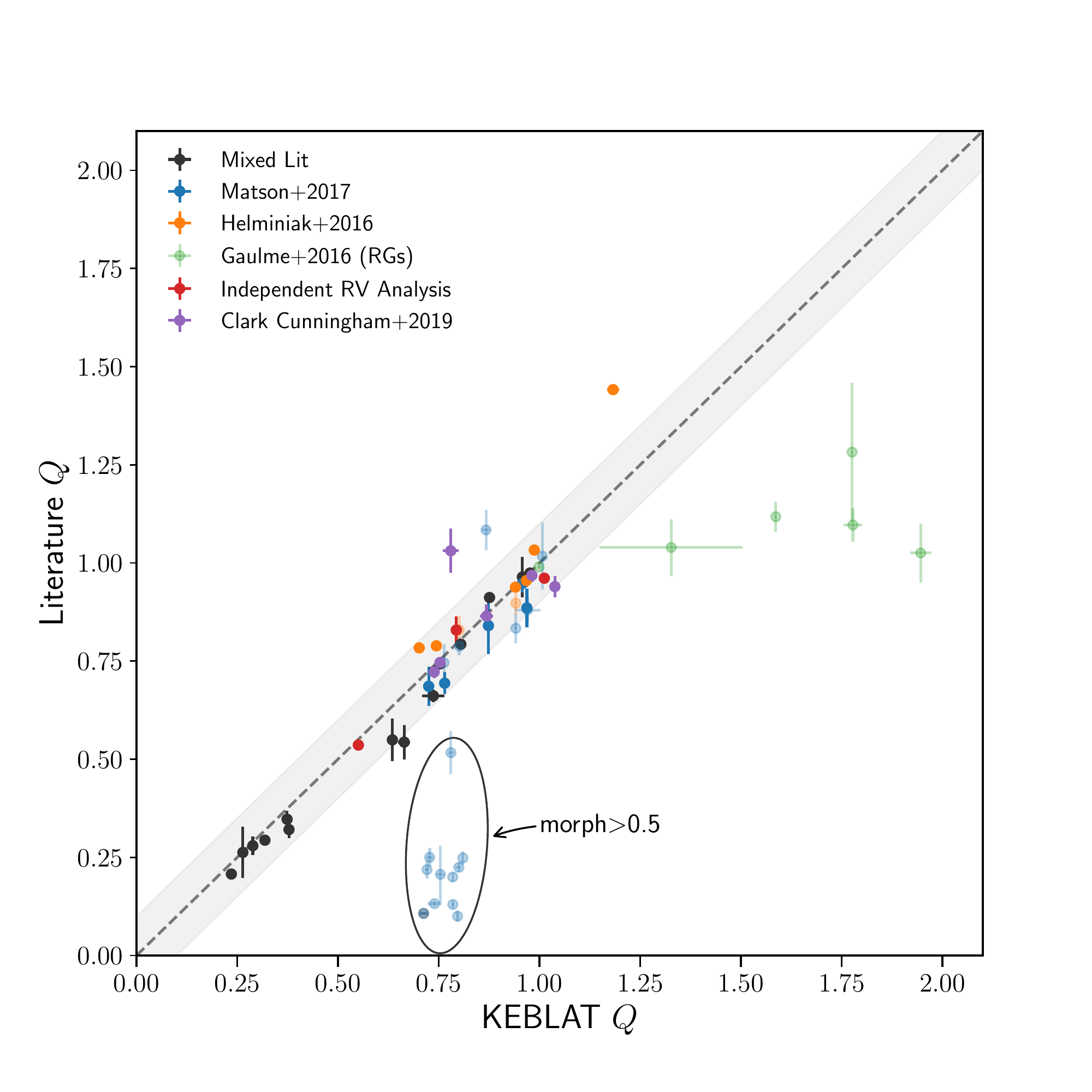}
\caption[width=0.5\textwidth]{Comparison of RV-derived mass ratio values as reported in literature vs. KEBLAT values using the SED+LC fitting method. The dashed black line denotes 1:1 relationship, with light grey regions representing $\pm$0.1 uncertainty in $Q$. In general, there is broad agreement with literature values. Binaries which are not well described by stellar isochrones (represented by lighter alpha values), e.g., red giants and binaries that exchange(d) mass, give discrepant mass ratio values.}
\label{fig:ebQcomparison} 
\end{figure*}

\begin{figure*}
\includegraphics[width=1.0\textwidth]{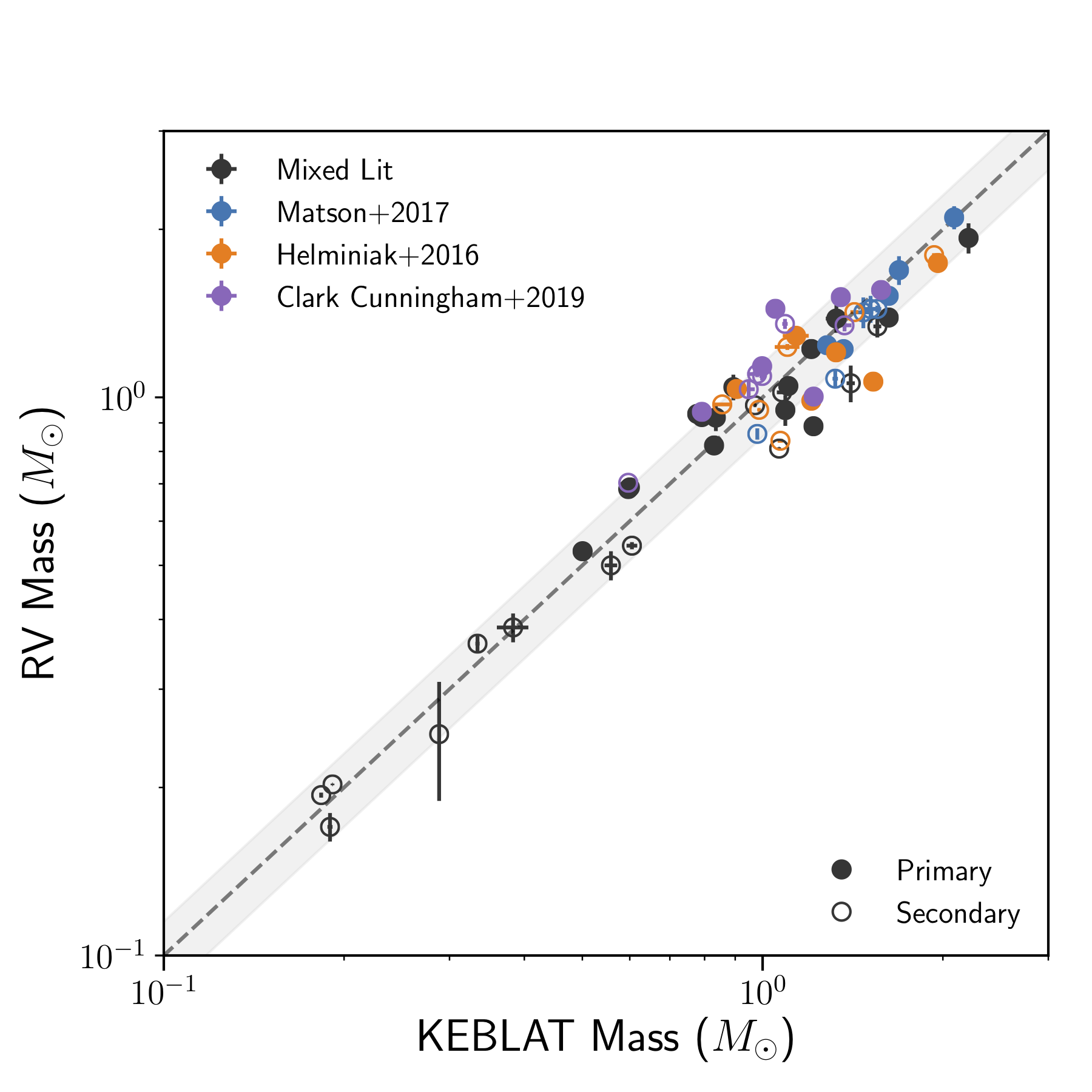}
\caption[width=0.5\textwidth]{Comparison of absolute mass values between KEBLAT and RV-derived values from literature, \edit{after removing EBs with red giant components and \texttt{morph}$>$0.5 (see \S\ref{subsec:mass})}. The closed and open circles represent primary and secondary components of the binary. The dashed black line denotes 1:1 relationship, with light grey regions representing 15\% of the mass uncertainties. The photometric masses show good agreement with RV values.}
\label{fig:eb_absolute_mass} 
\end{figure*}

\begin{figure*}
\includegraphics[width=0.9\textwidth]{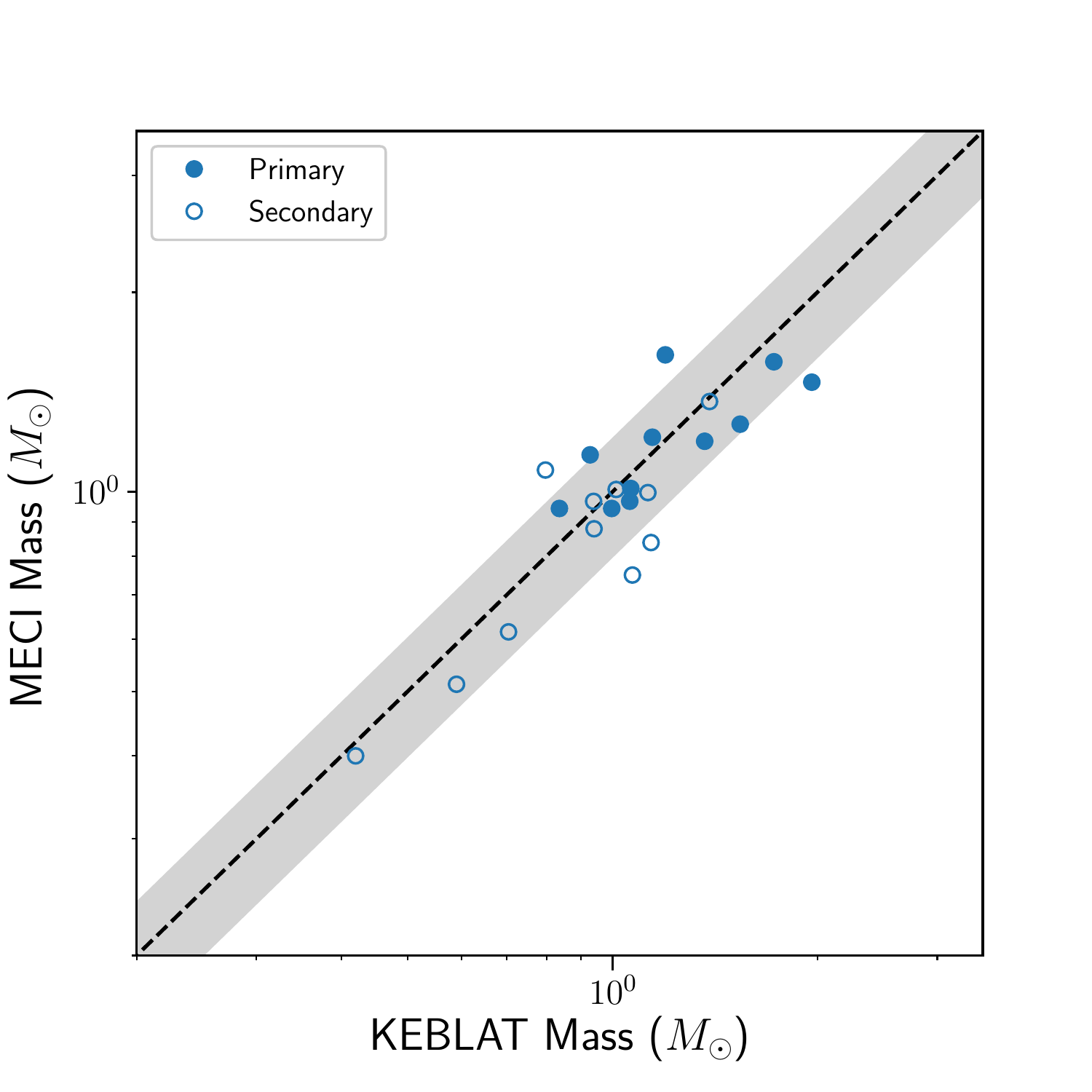}
\caption[width=0.5\textwidth]{Comparison of mass values from our analysis and that of \cite{Devor2008} for a small number of overlap binaries. Both studies used light curve and SED data with stellar isochrones to derive masses, although predicated on different data and details of each method are different. The mass values show relatively good agreement.}
\label{fig:Q_meci} 
\end{figure*}

\begin{figure*}
\includegraphics[width=0.9\textwidth]{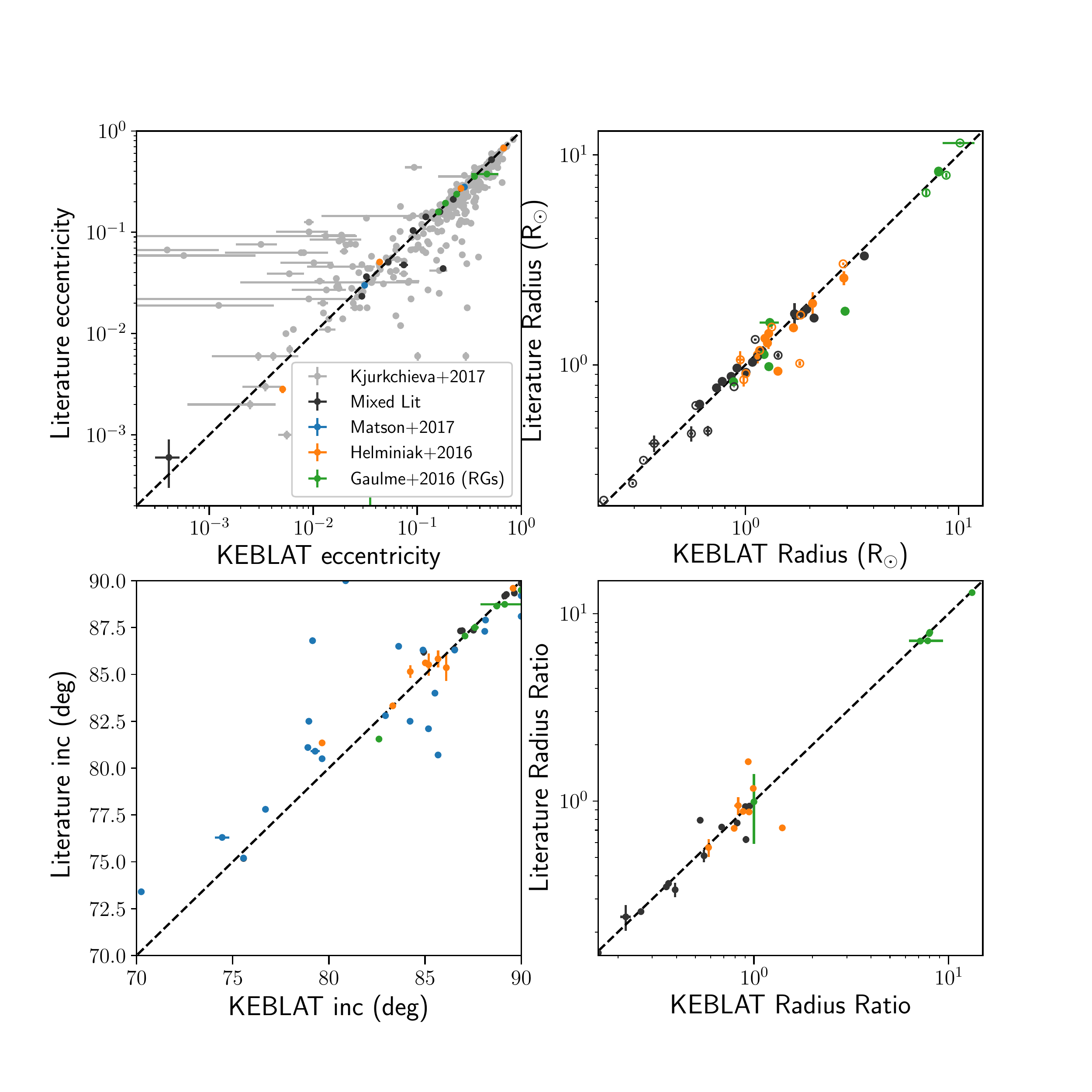}
\caption[width=0.5\textwidth]{Comparison of eccentricity, inclination, and radii values from KEBLAT and various literature studies, which show overall good agreement. The values are collected from the same RV studies from mass comparison, plotted in Fig.~\ref{fig:ebQcomparison}, and we supplement additional eccentricity estimates from \citep{Kjurkchieva2017}. While the masses of red giant components \citep{Gaulme2016} were poorly inferred, the absolute radii show good fidelity to literature values. There is also greater scatter among inclination values from \cite{Matson2017} (blue), however these were fixed to be \cite{Slawson2011} neural network inference values based on phenomenological light curve modeling rather than physical model, which may explain some of the discrepancies.}
\label{fig:eb_compare_e_r} 
\end{figure*}

\begin{figure*}
\includegraphics[width=\textwidth]{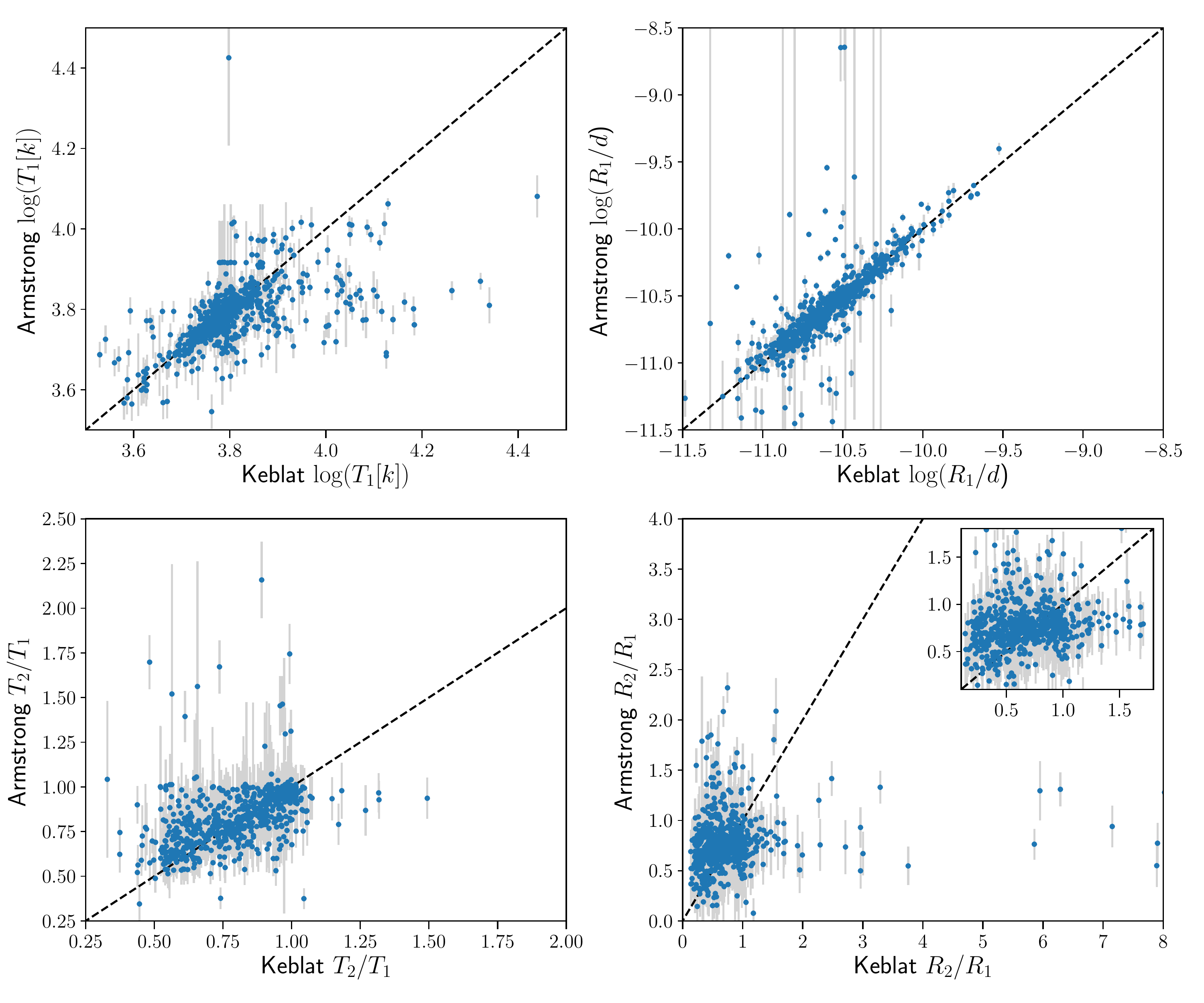}
\caption[width=0.5\textwidth]{Comparison of parameters as derived in this paper and values presented in \cite{Armstrong2014} for temperature of the primary star (top left), primary radius normalized by distance to system (top right), temperature ratio (bottom left), and radius ratio (bottom right). While $T_1$ and $R_1/d$ show bulk agreement, their temperature and radius ratios estimates are relatively crude and show significant scatter with respect to values derived in this work (see text for discussion).}
\label{fig:EB_Armstrong} 
\end{figure*}

In addition to stochastic ``white'' noise, \kepler light curves exhibit correlated noise, due to both instrumental systematics, e.g., telescope drift, and astrophysical effects, e.g., stellar variability, contaminant light from nearby sources (see \cite{Gilliland2015} for a discussion of \kepler\ noise properties); \edit{sources of astrophysical ``noise" may be equivalently considered as ``signals."} A low-order, local polynomial fitting around each eclipse is typically sufficient for light curves \edit{of detached systems} and minimal stellar variability. For EBs with strong quasi-periodic out of eclipse variation, such as those with \texttt{morph}$>0.5$ which comprise the majority of poor fits in Figure~\ref{fig:eb_medianresiduals}, a low-order polynomial may not sufficiently capture the variability, giving rise to larger residuals.

For single star targets, stellar variability, typically in the form of starspot rotation \citep{Giles2017}, stellar oscillation and granulation \citep{Bastien2013, North2017}, flares \citep{Davenport2016}, confound astrophysical signals. For EBs in our sample, the presence of a secondary star adds two levels of confusion: 1) stellar variability associated with the companion itself and 2) light curve effects due to dynamical interactions and binary evolution, such as ellipsoidal variations, gravity darkening, Doppler beaming, and reflection \citep{Faigler2013}. For case 2, the amplitudes of Doppler beaming and reflection/ellipsoidal effects for similar-mass, P$\lesssim$5 d solar-type binaries can be of order 1 and 10 ppt, respectively \citep{Zucker2007}. Binaries in our sample with morphology parameter $\gtrsim$ 0.5 are most affected by these effects, in particular ellipsoidal variations, and account for some of the population with larger residuals in Figure~\ref{fig:eb_medianresiduals} (see orange vs. blue bins). 

\edit{For the goal of large ensemble binary parameter inference from only photometry using MCMC, including a full physical treatment of tidal distortions and other photodynamical effects would be significantly more computationally expensive. Well-established numerical codes that include detailed physics already exist, such as \texttt{ELC} \citep{Orosz2000} and the \cite{Wilson1971} based code \texttt{PHOEBE} \citep{Prsa2016}, where stellar surfaces are modeled on a mesh or grid and allowed to deform. These numerical codes provide higher fidelity to the physics of close binaries at the expense of computational speed. A more computationally tractable method may be to use analytic approximations to explicitly describe ellipsoidal, beaming, and reflection effects, such as the \texttt{BEER} algorithm \citep{Faigler2011}. However, this approach would still benefit from an appropriate treatment of additional correlated instrumental noise and stellar variability.}

In principle, using Gaussian process (GP) kernels to model instrumental and astrophysical noise in conjunction with a simple physical binary model may lead to better light curve fitting fidelity. The kernel parameters may additionally give insight into the physical processes that underlie the observed quasi-periodic variations. For example, \cite{Angus2018} applied GPs to infer stellar rotation periods for a subset of \kepler objects of interests. In practice, however, we found that including GPs in our method to model \kepler EB light curves significantly increased computational time and required more user input on specific systems to assess kernel type and initial parameter choice; using GPs on \emph{K2} or \emph{TESS} light curves, which typically has $\sim$month long coverage rather than $\sim$year, may be more computationally tractable. We leave implementing GPs on correlated noise as future work, and note that \texttt{celerite}, a newly developed algorithm that scales with the number of data points $\mathcal{O}(N)$, might make significant improvements to computation time and fitting flexibility. 

\subsubsection{\edit{Third Light Contamination}}
\label{subsec:sed_noise}

Here, we consider the effects of ``third light contamination," or photometric contamination associated with background/foreground source(s) coincident to or companion(s) gravitationally bound to the binary. As mentioned in \S\ref{subsec:lightcurve} and \S\ref{subsec:sed}, we fit for quarterly \kepler contamination values during model optimization; however, we do not explicitly and self-consistently model flux dilution in the SED by additional sources, due to the limited number of SED measurements, as well as increased model complexity associated with specifying a third isochrone and differential point spread functions (PSFs) with wavelength. Instead, the SED module fits for a single systematic error term, i.e., any additional error beyond observational uncertainties, including third light contamination. This systematic error term should sufficiently capture contributions to the SED from relatively faint contaminators. 

\kepler photometry is more susceptible to flux contamination than the Johnson/SDSS/2MASS bands because of the instrument's large pixel scale and aperture radius\footnote{https://keplerscience.arc.nasa.gov/the-kepler-space-telescope.html} \cite[e.g.,][]{Schwamb2013, Morton2012}. While WISE bands 1 and 2 share similarly large aperture radii ($\sim$6\arcsec; \citealp{Kennedy2012}), the PSFs from optical and near-IR observations are small. As a result, many of the sources that apppear blended in \kepler and WISE photometry are expected to be resolved and contribute minimally to the bulk of the SED data in the optical and near-IR. 

Unresolved contaminators (e.g., point sources at small angular separations $<$1\arcsec) that are bright relative to the binary (e.g., giants or similar mass dwarfs) will introduce wavelength- dependent bias to the observed SED. If their contribution deviates significantly from \kepler quarterly estimates, they will also dilute the expected depth of eclipses. As we explain below, overall, we expect that only a small fraction (few percent) of our sample are deleteriously affected by bright, unresolved contaminants.

We estimate the extent to which these sources may affect our sample in two ways: 1) by using order of magnitude statistical arguments and 2) by identifying correlated characteristics in inferred model parameters. In the former case, we make the assumption that the majority of these unresolved sources are gravitationally bound, co-eval tertiaries. \cite{Tokovinin2006} found that the fraction of SB2s with tertiary companions in the period range $P_{\mathrm{tertiary}}=2-10^5$ yr depends strongly on the inner binary period, reaching $\sim$90\% for $P\lesssim3d$ and $\sim$35\% for $P\gtrsim$10 d. Studies of triples around largely short period ($P<10$ d) \kepler EBs, with $P_{\mathrm{tertiary}}\lesssim4$ yr, have converged on an occurrence rate of $\sim$15\% \citep[e.g.,][]{Gies2012, Rappaport2013, Conroy2014}. Based on these values, we assume that $\sim$30\% of our EBs host additional unresolved companions. Since the majority of tertiaries have detected masses smaller than the binary mass \citep{Tokovinin2008, Borkovits2015}, we estimate that 1/3 of triples, or $\sim$10\% of EBs in our sample may harbour companions with significant flux contribution. 

In our analysis, we expect that systems with substantial wavelength dependent flux contamination will introduce bias to the total spectral shape (skew) and to the distance modulus (offset). As a result, they may exhibit large inferred systematic SED errors $\sigma_{\mathrm{sys, SED}}$ and distance errors $\sigma_d$. We find that 5\% of our sample have $\sigma_{\mathrm{sys, SED}}>0.1$ mag and $\sigma_d>200$ pc. This value of a few percent appears consistent with the order of magnitude estimate from the statistical approach. However, as discussed in \S\ref{subsec:sed}, \S\ref{subsec:noise}, and \S\ref{subsec:age}, other effects can also cause large $\sigma_d$ and $\sigma_{\mathrm{sys, SED}}$ values. 

As an example, we note that KIC 4862625 is a binary with a pair of companions separated by 0.7\arcsec\ in a hierarchical quadruple configuration. The companion binary contributes $\sim$10\% and $\sim$20\% of the total flux in the \kepler and $K_s$ bands \citep{Schwamb2013}. Despite this, our stellar inferences ($M_1=1.33$\msun, $M_2=0.38$\msun) agree remarkably well with published values ($M_1=1.38-1.53$\msun, $M_2=0.387-0.408$\msun; \citealp{Schwamb2013, Kostov2013}). 

\subsubsection{Isochrone inaccuracies \& fitting biases}
\label{subsec:age}
An inherent limitation to our model is that it relies on theoretical stellar models, which may contain inaccuracies and generate fitting biases. In particular, the complex morphology of isochrones challenges uniform parameter space sampling \citep{Jorgensen2005}, and the predictive power of isochrones is weak for low mass dwarfs (the radius inflation problem; \citealp{Stassun2012}), post-MS evolutionary phases (e.g., red giants; \citealp{Cassisi2017}), and stars that exchange(d) mass. These factors lead to discrepant mass values when compared to RV studies (\S\ref{subsec:mass}) and the population of over-represented young ($\tau<8$) stars in our general sample (\S\ref{subsec:stellar}). While young ages may be used as consistency checks to the fidelity of stellar isochrones, stringent tests and calibration of stellar evolution models will require the bulk collection of precise radial-velocity measurements of these eclipsing binaries, such as expanding upon the initial survey of EBs with APOGEE \citep{Zasowski2013,Badenes2018}.  

The morphology of stellar isochrones, in which small perturbations in mass, age, or metallicity can lead to large differences in radius, temperature, and luminosity, challenges uniform sampling of input parameters based on observables. This is particularly true during post-MS evolutionary phases, which apply to RG binaries in our sample. While masses do not always agree, KEBLAT radius values track RV solutions remarkably well for them (see Figure~\ref{fig:eb_compare_e_r}). This indicates that the isochrones are converging to young, higher mass MS secondaries with large radius values in order to reconcile $R_2/R_1\gg1$ based on the eclipses. One means of improving isochrone interpolation for future implementations may be to adopt an ``Equivalent Evolutionary Phases" method \citep{Dotter2016}, in which isochrones are transformed onto a uniform basis that samples evenly different evolutionary phases, rather than evenly in absolute timesteps. Another potential solution for better sampling is to fit the isochrones with a multi-dimensional Gaussian process and compute derivatives of the GP with respect to model parameters; these derivatives would enable more robust optimization and Hamiltonian Monte Carlo analysis to more efficiently sample parameter space.

Another reason why isochrone fitting might fail for red giants is that theoretical radius predictions may be too uncertain (e.g., poor constraints on mass loss, convective overshooting parameters) for the level of precision needed to fit flat-bottomed eclipses in the \kepler light curves. Astereoseismic studies of red giants, which have been revolutionized in the era of \kepler and \emph{CoRoT}, can provide more reliable and physically motivated constraints on mass relative to our method, and better calibrate stellar models.

In addition to post-MS evolution uncertainty, theoretical models do not reproduce observations of M-dwarfs well. Several studies have shown that low-mass ($M\approx0.1-0.8$\msun) dwarfs have radii inflated relative to theoretical models, up to $\sim15$\% \citep{Stassun2012, Kraus2011, Cruz2018}. This radius inflation is coupled by a corresponding decrease in effective temperature, such that the mass-luminosity relationship (for MS stars) is preserved. The presence of radius discrepancies appears to be agnostic to stellar multiplicity \citep{Kesseli2018}. Multiple studies agree that radius inflation is associated with magnetic field activity, although they diverge in their conclusions about the relative importance of rotation-induced surface magnetic fields and their potential effects on stellar radii via convective inhibition, dark spot coverage, and tidal synchronization in binaries \citep{Chabrier2007, Stassun2012, Feiden2014, Kesseli2018}. 

The PARSEC isochrones we used in our SED+LC model included improved calibration for low mass stars with radius discrepancy at the $\sim5$\% level \citep{Chen2014}. Binaries with anomalously young M-dwarfs may be explained by the model attempting to compensate for a larger radius than theoretical model predicts; stars that have not yet started core H fusion, i.e., reached the main sequence, are still contracting, and thus have larger radii compared to their MS counterparts. For M dwarfs, the pre-MS stage may last $\sim10$ Myr to $\sim1$ Gyr, consistent with inferred youthful ages. Allowing a factor for radius inflation might help in modeling these stars. However, because the luminosity-mass relation for M dwarfs is well calibrated, a radius inflation factor necessarily requires a corresponding scaling relation to lower the effective temperature, which in turn affects the predicted SED. This makes treating the issue in a self-consistent way difficult.

Finally, our model assumes that stars fall on coeval isochrones, an assumption which can be strongly violated in binaries which have undergone mass-transfer. The majority of binaries with anomalously young ages in our general sample and with discrepant masses in our RV comparison sample appear to be Algol-type binaries. These are systems with primary masses between 2-4 \msun  \citep{Ibanoglu2006}, V-shaped eclipses, and typically exhibiting semi-detached morphologies (see e.g., Figure~\ref{fig:10619109}). In these eclipsing Algol systems, mass transfer occurs from the more evolved and initially more massive component which has filled its Roche lobe onto the less massive companion; as the companion accretes more mass, it is ``rejuvenated" becoming hotter, more massive, and delaying evolution away from main sequence. Stellar evolutionary tracks of single-stars cannot accurately capture masses and ages of stars in Algol-type binaries because they do not take into account modification due to binary evolution. Because the primary components of Algol-type binaries are typically ``main sequence" stars with high mass (O, B, A dwarfs), we expect the ages of this population to be skewed toward young populations. One possible ad-hoc solution to these rejuvenated systems is to allow the ages of both stars to vary independently, although such a treatment may not be physically self-consistent\edit{, and the additional degree of freedom may pose further difficulties in efficiently sampling the full parameter space}. Alternatively, the same mechanism that enables mass transfer, i.e., when a star fills its Roche lobe, can be used to constrain mass ratios in EBs with semi-detached or over-contact light curve morphologies; this method has demonstrated success for totaling eclipsing SD or OC systems \citep{Wyithe2002}.

\section{Conclusions}
\label{sec:conclusions}

We have modeled the light curves and SEDs of 728 $\sim$detached \kepler eclipsing binaries, using Bayesian inference for full binary parameter estimation. Our forward model couples Padova stellar isochrones to a \cite{Mandel2002} transit model to solve the system's stellar and orbital parameters with only photometric information. The entire set of posteriors and maximum likelihood solutions are available online.

We are interested in obtaining reliable orbital and mass solutions, in particular as the basis of an automated and optimized search for circumbinary planets. We find that binary orbital parameters are typically very well constrained by the light curve data, and comparisons for a subset of our sample with previous studies indicate that the majority are reliable. 

Moreover, we show that detached, main sequence binaries give accurate masses relative to RV ``truths." Binary systems which have been affected by stellar evolution, or contain stars which are otherwise discrepant with isochrone predictions, however, may give poor agreement. The latter exhibit symptoms of higher morphology values (\texttt{morph}$\gtrsim$0.5) and artificially young stellar ages ($\tau\lesssim7.5$), which may be used as a filter for quality check. These symptoms are manifestations of the inherent limitations associated with our approach: namely, that we do not account for non-Keplerian effects like ellipsoidal variations, and that we rely on co-eval isochrone models; the inferred stellar parameters are only as accurate as input theoretical models. On the other hand, discrepant results that arise from these limitations may help inform which binary targets warrant more time-intensive follow-up observations, to better calibrate theoretical stellar models. For future work, our methodology may be improved by additional \emph{Gaia} systemic radial velocities. Currently \emph{Gaia} does not provide these for targets flagged as binaries; once \emph{Gaia} releases its spectroscopic catalogue, systemic RV can be estimated and, in conjunction with proper motion, the full 3-dimensional motion can provide additional constraints to binary ages based on galactic kinematics. Additionally, as we have discussed in \S\ref{subsec:limitations}, incorporating Gaussian processes into the time series and isochrone modeling may improve the robustness of our model to non-Keplerian light curve effects and post-MS stars. 

We conclude by summarizing our key takeaways below:
\begin{enumerate}
    \item We have created the largest, homogeneous catalogue of \kepler eclipsing binaries with full system parameters and posteriors from Bayesian inference. This catalogue is of interest to identify compelling targets for RV follow-up (e.g., M-dwarf binaries), and to binary population studies which require mass estimates and precise measurements of orbital elements (e.g., eccentricities), such as tidal theory and circumbinary planet searches. 
    \item We additionally identify 35 new systems with eclipse timing variations, perhaps arising from apsidal motion due to binary tides and/or tertiary companions.  
    \item We demonstrate that we can reliably retrieve mass using only photometry for well-detached, main sequence binaries, using comparisons to RV ``truths." These systems make up the majority of our sample. 
    \item For a subset of binaries with post-main sequence components or significant tidal deformations (e.g., semi-detached systems where mass exchange have occurred), our inferred masses are not universally reliable, as our model does not account for non-spherical distortions and are limited by the fidelity of input stellar isochrones. We recommend using $\tau\lesssim7.5$ and \texttt{morph}$>$0.5 to diagnose potentially problematic systems. Fortunately, photometric masses may be derived using asteroseismology \citep{Gaulme2016} for red giants and by exploiting Roche lobe-filling configurations for semi-detached and over-contact binaries \citep{Wyithe2002}. 
    \item The technique used here can be applied to OGLE, TESS, and other large time-domain photometric surveys coming online, such as LSST, to characterize large numbers of eclipsing binaries. As many of these surveys have all-sky coverage, this poses an intriguing opportunity to probe binary galactic archeaology. 
\end{enumerate}

\section{acknowledgements}
This work was supported by NASA grants NNX12AF20G, NNX13AF62G, NNX15AT44H (NESSF), and NSF grant 1615315. This work was facilitated through the use of advanced computational, storage, and networking infrastructure provided by the Hyak supercomputer system at the University of Washington, and made use of the gaia-kepler.fun crossmatch database created by Megan Bedell. We thank the anonymous referee for their helpful inputs and constructive criticism. DW would like to thank Dr. Leslie Hebb and Dr. James Davenport for useful discussions on binaries. 

\clearpage

\begin{deluxetable}{llllll}
\tabletypesize{\footnotesize}
\tablecolumns{6} 
\tablewidth{0pt}
 \tablecaption{ Model Parameters 
 \label{tab:modelparameters}}
 \tablehead{
 \colhead{Parameter} & \colhead{Meaning} & \colhead{Module$^{\dagger}$} & \colhead{Constrained by} & \colhead{Priors}  \\ 
 } 
 \startdata  
    $P$                          & period (d)                                                                          & LC          & time of eclipses               & flat (0, 4000)     \\
    $e\sin\omega$                & transformation of eccentricity                    & LC          & ratio of PE + SE durations        & flat (-0.95, 0.95) \\
    $e\cos\omega$                & \ \ \ \  \& longitude of periastron                    & LC          & time of PE + SE                & flat (-0.95, 0.95) \\
    $t_{\mathrm{PE}}$            & time of primary eclipse (BJD-2454833)                                                         & LC          & LC minimum                                    & flat (0, 4000)     \\
    $b$                          & impact parameter $\frac{a \cos i}{ R_1}$                                                   & LC          & shape of eclipse                  & flat (-10, 10)     \\
    $q_{11}, q_{12}$                   & transformed quadratic limb darkening  & LC  & shape of eclipse                  & flat [0, 1]        \\
    $q_{21}, q_{22}$                   & \ \ \ coeffs for primary \& secondary & LC          & shape of eclipse                  & flat [0, 1]        \\
    $F_{\mathrm{rat}}$           & Kepler flux ratio $\frac{F_2}{F_1}$                                             & LC only                & relative depth of eclipse         & flat [1e-8, 100]   \\
    $R_{\mathrm{sum}}$           & sum of radii (\rsun)                                                                & LC only                & duration of eclipse             & flat [0.1, 1e6]    \\
    $R_{\mathrm{rat}}$           & ratio of radii $\frac{R_2}{R_1}$                                                            & LC only                & eclipse shape + relative depth  & flat [1e-6, 1000]  \\
    $M_{\mathrm{sum}}$           & sum of masses (\msun)                                                               & SED, LC*    & shape of SED                          & flat [0.2, 24]     \\
    $Q$          & ratio of masses $\frac{M_2}{M_1}$                                                           & SED         & shape of SED                          & flat [0.0085, 2]   \\
    $z$                        & binary metallicity, defined as 1-X-Y ($z_{\odot}=0.015$)                                                                & SED         & shape of SED                          & flat [0.001, 0.06] \\
    $\tau$                       & binary $\log_{10}$ age ($\log_{10}$(yr))                                                            & SED         & shape of SED                          & flat [6, 10.1]     \\ 
    $d$                          & distance to system (pc)                                                             & SED         & SED offset                            & Gaussian${\ddagger}$  $d_{\mathrm{Gaia}}, \sigma_d$  \\
    $E(B-V)$                     & reddening, assuming $R_V=3.1$                                         & SED         & shape of SED                          & Gaussian $\mathrm{E(B-V)}_{\mathrm{Schlafly}}, \sigma_{\mathrm{E(B-V)}}$         \\
    $h_0$                        & dust vertical scaleheight (pc)                                        & SED         & shape of SED                          & fixed at 119            \\
    $\sigma_{d}$ & uncertainty in \emph{Gaia} distance prior  ($\ln$)                                                       & SED         & ~                                     & flat [-1, 7]     \\
    $\sigma_{\mathrm{E(B-V)}}$ & uncertainty in reddening prior  ($\ln$)                                                       & SED         & ~                                     & flat [-12, 2]     \\
    $\sigma_{\mathrm{sys, SED}}$ & systematic SED error  ($\ln$)                                                       & SED         & ~                                     & flat [-25, 0]     \\
    $\sigma_{\mathrm{sys, LC}}$  & systematic LC error  ($\ln$)                                                        & LC          & ~                                     & flat [-25, -4]     \\ 
    $c_{j}$  & crowding parameter for quarter $j$                                                        & LC          & depth of eclipses                                     & flat [0, 1]     \\ 
\enddata
\vspace{-0.8cm}
\tablecomments{${\dagger}$All parameters belong to the simultaneous SED+LC module, unless otherwise specified, e.g., ``LC only."  \\ *When only light curve data available, $M_{\mathrm{sum}}$ is a proxy for $a$, to obtain the sizes of stars relative to their separation. \\ ${\ddagger}$ When no \emph{Gaia} parallax available, the distance prior is flat on [10, 15000] pc.}
\end{deluxetable}

\begin{deluxetable}{lllllll}
\tabletypesize{\footnotesize}
\tablecolumns{7} 
\tablewidth{0pt}
 \tablecaption{ Eclipsing Binary Orbital \& Distance Parameter Posteriors 
 \label{tab:ebresults1}}
 \tablehead{
 \colhead{KIC} & \colhead{$P$} & \colhead{$t_{\mathrm{PE}}$} & \colhead{$e\sin\omega$} & \colhead{$e\cos\omega$} 
 & \colhead{$i$} & \colhead{$d$} \\ 
 ~ & \colhead{(d)} & \colhead{(BJD-2454833)} & ~ & ~ & \colhead{ (rad)} & \colhead{(pc)}
 } 
 \startdata  
10031409	&	4.143879363	$^{+	4.5e\text{-}8	}_{	-4.6e\text{-}8	}$ &	124.5491837	$^{+	9.6e\text{-}6	}_{	-1.0e\text{-}5	}$ &	0.0047	$^{+	0.0014	}_{	.0012	}$ &	0.000007	$^{+	0.000004	}_{	.000004	}$ &	1.4923	$^{+	0.0002	}_{	.0002	}$ &	941	$^{+	23	}_{	-22	}$ \\
10198109	&	17.91874241	$^{+	3.8e\text{-}7	}_{	-3.9e\text{-}7	}$ &	146.3930259	$^{+	1.8e\text{-}5	}_{	-1.8e\text{-}5	}$ &	0.2787	$^{+	0.0025	}_{	.0024	}$ &	0.2484	$^{+	0.0002	}_{	.0002	}$ &	1.5825	$^{+	0.0003	}_{	.0003	}$ &	637	$^{+	48	}_{	-22	}$ \\
12356914	&	27.30822669	$^{+	9.3e\text{-}6	}_{	-9.4e\text{-}6	}$ &	143.5001244	$^{+	2.7e\text{-}4	}_{	-2.8e\text{-}4	}$ &	0.4245	$^{+	0.0154	}_{	.0153	}$ &	0.1213	$^{+	0.0009	}_{	.0010	}$ &	1.5642	$^{+	0.0014	}_{	.0013	}$ &	1344	$^{+	1	}_{	-14	}$ \\
10619109	&	2.045166167	$^{+	3.4e\text{-}8	}_{	-4.7e\text{-}8	}$ &	122.139347	$^{+	1.7e\text{-}5	}_{	-1.6e\text{-}5	}$ &	0.0676	$^{+	0.0043	}_{	.0043	}$ &	0.00136	$^{+	0.00008	}_{	.00013	}$ &	1.2260	$^{+	0.0027	}_{	.0027	}$ &	1081	$^{+	24	}_{	-22	}$ \\
12644769	&	41.07759097	$^{+	1.4e\text{-}6	}_{	-1.4e\text{-}6	}$ &	132.6583057	$^{+	2.8e\text{-}5	}_{	-2.9e\text{-}5	}$ &	.1641	$^{+	0.0005	}_{	.0005	}$ &	.017971	$^{+	0.000004	}_{	.000004	}$ &	1.57659	$^{+	0.00002	}_{	.00001	}$ &	68.6	$^{+	0.5	}_{	.5	}$ \\
\enddata
\vspace{-0.8cm}
\tablecomments{Posteriors for orbital \& distance parameters for a subset of our sample which corresponds to Figures~\ref{fig:10031409} to \ref{fig:12644769}, to demonstrate the table's form and content. Full posteriors for all parameters available in online table. See Table~\ref{tab:modelparameters} for details on parameter meanings.}
\end{deluxetable}

\begin{deluxetable}{lllllllll}
\tabletypesize{\footnotesize}
\tablecolumns{9} 
\tablewidth{0pt}
 \tablecaption{ Eclipsing Binary Stellar Parameter Posteriors 
 \label{tab:ebresults2}}
 \tablehead{
 \colhead{KIC} & \colhead{$z$} & \colhead{$\tau$} & \colhead{$M_1$} & \colhead{$M_2$} 
 & \colhead{$R_1$} & \colhead{$R_2$} & \colhead{$F_2/F_1$} \\
 ~ & ~  & \colhead{(log(yr))} & \colhead{$(M_{\odot})$} & \colhead{$(M_{\odot})$} & \colhead{$(R_{\odot})$} & \colhead{$(R_{\odot})$} & ~ 
 } 
 \startdata  
10031409	&	0.011	$^{+	0.002	}_{	-0.002	}$ &	9.37	$^{+	0.16	}_{	-0.20	}$ &	1.13	$^{+	0.04	}_{	-0.05	}$ &	1.08	$^{+	0.04	}_{	-0.05	}$ &	1.183	$^{+	0.009	}_{	-0.009	}$ &	1.085	$^{+	0.022	}_{	-0.028	}$ &	0.7689	$^{+	0.0245	}_{	-0.0289	}$ \\
10198109	&	0.006	$^{+	0.001	}_{	-0.002	}$ &	9.57	$^{+	0.28	}_{	-0.71	}$ &	1.01	$^{+	0.15	}_{	-0.08	}$ &	0.36	$^{+	0.03	}_{	-0.01	}$ &	1.078	$^{+	0.045	}_{	-0.025	}$ &	0.347	$^{+	0.014	}_{	-0.008	}$ &	0.00473	$^{+	0.00002	}_{	-0.00002	}$ \\
12356914	&	0.007	$^{+	0.002	}_{	-0.001	}$ &	9.77	$^{+	0.20	}_{	-0.33	}$ &	0.94	$^{+	0.08	}_{	-0.06	}$ &	0.60	$^{+	0.02	}_{	-0.02	}$ &	1.026	$^{+	0.014	}_{	-0.018	}$ &	0.587	$^{+	0.014	}_{	-0.015	}$ &	0.0602	$^{+	0.0020	}_{	-0.0013	}$ \\
10619109	&	0.0023	$^{+	0.0003	}_{	-0.0002	}$ &	6.25	$^{+	0.01	}_{	-0.01	}$ &	2.48	$^{+	0.04	}_{	-0.05	}$ &	1.87	$^{+	0.02	}_{	-0.02	}$ &	2.619	$^{+	0.016	}_{	-0.020	}$ &	3.038	$^{+	0.018	}_{	-0.018	}$ &	0.2093	$^{+	0.0131	}_{	-0.0118	}$ \\
12644769	&	0.0100	$^{+	0.00005	}_{	-0.00004	}$ &	10.0796	$^{+	0.0003	}_{	-0.0006	}$ &	0.6002	$^{+	0.0005	}_{	-0.0003	}$ &	0.1913	$^{+	0.0003	}_{	-0.0002	}$ &	0.6092	$^{+	0.0004	}_{	-0.0002	}$ &	0.2161	$^{+	0.0003	}_{	-0.0002	}$ &	0.01582	$^{+	0.00003	}_{	-0.00003	}$ \\
\enddata
\vspace{-0.8cm}
\tablecomments{Posteriors for stellar parameters for a subset of our sample which corresponds to Figures~\ref{fig:10031409} to \ref{fig:12644769}, to demonstrate the table's form and content. Full posteriors for all parameters available in online table. See Table~\ref{tab:modelparameters} for details on parameter meanings.}
\end{deluxetable}

\begin{deluxetable}{cccccc}
\tabletypesize{\footnotesize}
\tablecolumns{6} 
\tablewidth{0pt}
 \tablecaption{ Maximum Likelihood Parameter Solutions
 \label{tab:ebresults_ml}}
 \tablehead{
 ~ & \colhead{KIC} &  \colhead{KIC} & \colhead{KIC} &  \colhead{KIC} & \colhead{KIC} \\
 \colhead{Parameter} & \colhead{10031409} & \colhead{10198109} & \colhead{12356914} & \colhead{10619109} & 
 \colhead{12644769}
 } 
 
 \startdata  

$M_{\mathrm{sum}}$ &	2.2751	&	1.2485	&	1.5222	&	4.3590	&	0.7919	\\
$Q$ &	0.9536	&	0.3731	&	0.6604	&	0.7520	&	0.3188	\\
$z$ &	0.0138	&	0.0079	&	0.0087	&	0.0023	&	0.0100	\\
$\tau$ &	9.2752	&	9.9106	&	9.8953	&	6.2404	&	10.0800	\\
$d$ &	963.48	&	608.77	&	1343.87	&	1073.79	&	68.63	\\
E(B-V) &	0.2117	&	0.2395	&	0.2214	&	0.5907	&	0.0664	\\
$h_0$ &	119	&	119	&	119	&	119	&	119	\\
$P$ &	4.143879379	&	17.9187423	&	27.30822721	&	2.04516617	&	41.07759104	\\
$t_{\mathrm{PE}}$ &	124.549181	&	146.3930288	&	143.5001619	&	122.1393461	&	132.6583081	\\
$e\sin\omega$ &	0.003968	&	0.276679	&	0.443474	&	0.066164	&	-0.163677	\\
$e\cos\omega$ &	0.000006	&	0.248540	&	0.120055	&	0.001243	&	-0.017973	\\
$b$ &	0.9400	&	-0.3405	&	0.3313	&	1.4283	&	-0.4410	\\
$q_{11}$ &	0.1732	&	0.2798	&	0.4472	&	0.9994	&	0.1763	\\
$q_{12}$ &	0.3858	&	0.4381	&	0.8051	&	0.0002	&	0.9990	\\
$q_{21}$ &	0.3624	&	0.6030	&	0.9237	&	0.5990	&	0.2659	\\
$q_{22}$ &	0.1984	&	0.9802	&	0.8589	&	0.3190	&	0.8639	\\
$\ln \sigma_{\mathrm{sys,LC}}$ &	-7.2	&	-8.1	&	-4.3	&	-6.5	&	-8.0	\\
$\ln \sigma_{\mathrm{sys,SED}}$ &	-3.5	&	-3.7	&	-3.0	&	-2.4	&	-3.4	\\
$\ln \sigma_{\mathrm{E(B-V)}}$ &	-11.1	&	-2.0	&	-2.0	&	-0.7	&	-6.4	\\
$\ln \sigma_{\mathrm{d}}$ &	0.0	&	6.7	&	-6.1	&	0.0	&	2.4	\\
$c_0$ &	0.9917	&	0.0000	&	0.0000	&	0.9885	&	0.9847	\\
$c_1$ &	0.9909	&	0.9927	&	0.9339	&	0.9885	&	0.9784	\\
$c_2$ &	0.9898	&	0.9909	&	0.9846	&	0.9901	&	0.9859	\\
$c_3$ &	0.9864	&	0.9910	&	0.9930	&	0.9897	&	0.9902	\\
$c_4$ &	0.9867	&	0.9902	&	0.9793	&	0.9886	&	0.9902	\\
$c_5$ &	0.9910	&	0.9927	&	0.9362	&	0.9884	&	0.9849	\\
$c_6$ &	0.9927	&	0.9925	&	0.9884	&	0.9897	&	0.9874	\\
$c_7$ &	0.0000	&	0.9910	&	0.9932	&	0.0000	&	0.9929	\\
$c_8$ &	0.9888	&	0.9902	&	0.9617	&	0.9885	&	0.9849	\\
$c_9$ &	0.9887	&	0.9927	&	0.8870	&	0.9884	&	0.9842	\\
$c_{10}$ &	0.9925	&	0.9925	&	0.9718	&	0.9897	&	0.9928	\\
$c_{11}$ &	0.0000	&	0.9909	&	0.9667	&	0.0000	&	0.9888	\\
$c_{12}$ &	0.9887	&	0.9892	&	0.9872	&	0.9885	&	0.9922	\\
$c_{13}$ &	0.9899	&	0.9927	&	0.9442	&	0.9885	&	0.9814	\\
$c_{14}$ &	0.9851	&	0.9925	&	0.9958	&	0.9897	&	0.9955	\\
$c_{15}$ &	0.0000	&	0.9909	&	1.0000	&	0.0000	&	0.9859	\\
$c_{16}$ &	0.9877	&	0.9892	&	0.9902	&	0.9885	&	0.9949	\\
$c_{17}$ &	0.9904	&	0.9927	&	0.9388	&	0.9885	&	0.9818	\\
\texttt{morph}$^{\dagger}$ &	0.35	&	0.07	&	0.03	&	0.55	&	0.03	\\

\enddata
\vspace{-0.8cm}
\tablecomments{Maximum likelihood solutions for a subset of our sample which corresponds to Figures~\ref{fig:10031409} to \ref{fig:12644769}, to demonstrate the table's form and content. Full version available online. See Table~\ref{tab:modelparameters} for details on parameter meanings and units. Values of 0.0 denotes no data was available. ${\dagger}$ \texttt{morph} is the morphology value taken directly from the Villanova \kepler Eclipsing Binary Catalogue \citep{Prsa2011, Kirk2016}. }
\end{deluxetable}

\begin{deluxetable}{cccccccccc}
\tabletypesize{\footnotesize}
\tablecolumns{7} 
\tablewidth{0pt}
 \tablecaption{ETV candidates identified in our sample
 \label{tab:etv_candidates}}
 \tablehead{
 \colhead{KIC} & \colhead{P (d)} & \colhead{Q} & \colhead{$e$} & \colhead{\edit{Other Sources}$^{\dagger}$}
 }
 \startdata  
10095512 & 6.01720794   & 0.83 & 2.0e-03 & brc \\
10215422 & 24.84708632  & 0.64 & 2.9e-01 &     \\
10268903 & 1.10397904   & 0.82 & 4.5e-03 & b   \\
10292238 & 143.11911400 & 0.86 & 6.2e-01 &     \\
10296163 & 9.29674544   & 0.52 & 3.8e-01 & b   \\
10352603 & 32.77898645  & 1.01 & 4.7e-01 &     \\
10549576 & 9.08935880   & 0.52 & 6.9e-02 & b   \\
10583181 & 2.69635500   & 0.58 & 3.2e-07 & b   \\
10619109 & 2.04516617   & 0.75 & 6.6e-02 &     \\
10686876 & 2.61841459   & 0.46 & 1.0e-03 & b   \\
10736223 & 1.10509420   & 0.72 & 1.2e-02 &     \\
10909274 & 39.23813527  & 1.00 & 7.3e-01 &     \\
10979716 & 10.68409498  & 0.80 & 1.5e-01 & b   \\
11071207 & 8.04963317   & 1.01 & 3.1e-01 &     \\
11234677 & 1.58742711   & 0.45 & 1.6e-01 & b   \\
11499757 & 12.31440178  & 0.89 & 2.6e-01 &     \\
11558882 & 73.91782410  & 0.60 & 4.2e-01 & b   \\
11724091 & 1.55909021   & 0.50 & 1.9e-02 &     \\
11923819 & 33.15943126  & 1.02 & 3.3e-01 &     \\
12356914 & 27.30822721  & 0.66 & 4.6e-01 & b   \\
12459731 & 14.18110967  & 0.77 & 4.1e-02 & b   \\
12644769 & 41.07759104  & 0.32 & 1.6e-01 &     \\
1995732  & 77.36197359  & 1.09 & 1.3e-01 &     \\
2305372  & 1.40469172   & 0.77 & 2.8e-02 & b   \\
2306740  & 10.30698967  & 0.86 & 3.1e-01 &     \\
2576692  & 87.87820528  & 1.00 & 2.1e-01 & b   \\
3241619  & 1.70334728   & 0.76 & 6.5e-03 &     \\
3247294  & 67.42012708  & 1.33 & 6.0e-01 & o   \\
3440230  & 2.88110019   & 0.72 & 1.0e-01 & b   \\
3757778  & 36.51436879  & 0.43 & 3.6e-01 &     \\
4077442  & 0.69284258   & 0.74 & 1.2e-01 &     \\
4544587  & 2.18911101   & 0.88 & 2.8e-01 &     \\
4753988  & 7.30445219   & 0.71 & 1.1e-02 & b   \\
4773155  & 25.70600971  & 0.92 & 4.4e-01 &     \\
4848423  & 3.00364489   & 0.96 & 2.5e-03 & b   \\
4948863  & 8.64355067   & 0.68 & 4.8e-03 & b   \\
5039441  & 2.15138428   & 0.52 & 2.8e-03 & brc \\
5113053  & 3.18509078   & 0.87 & 6.0e-05 &     \\
5269407  & 0.95887119   & 0.53 & 2.3e-04 & b   \\
5288543  & 3.45707832   & 0.80 & 6.4e-03 &     \\
5513861  & 1.51021130   & 0.95 & 2.2e-06 & b   \\
5553624  & 25.76208222  & 0.83 & 5.6e-01 &     \\
5632781  & 11.02520265  & 1.01 & 2.7e-01 &     \\
5731312  & 7.94636806   & 0.77 & 4.7e-01 & b   \\
5955321  & 11.63787579  & 0.98 & 4.8e-01 &     \\
5962716  & 1.80459191   & 0.71 & 2.8e-01 & b   \\
6029130  & 12.59165677  & 0.94 & 1.5e-02 &     \\
6042116  & 5.40715640   & 0.64 & 9.7e-02 &     \\
6449358  & 5.77679432   & 0.43 & 4.0e-04 &     \\
6464285  & 0.84365137   & 0.61 & 1.6e-02 &     \\
6525196  & 3.42059775   & 0.94 & 7.1e-05 & brc \\
6543674  & 2.39103076   & 0.97 & 4.5e-02 & b   \\
6545018  & 3.99145640   & 0.83 & 2.7e-03 & brc \\
6610219  & 11.30099291  & 1.01 & 2.1e-01 &     \\
6877673  & 36.75871004  & 0.81 & 2.0e-01 & b   \\
7021177  & 18.64532032  & 0.97 & 5.9e-01 &     \\
7025540  & 2.14821893   & 0.96 & 1.2e-02 &     \\
7137798  & 2.25353766   & 0.46 & 1.1e-01 &     \\
7177553  & 17.99645567  & 1.45 & 4.2e-01 & b   \\
7257373  & 10.46690066  & 0.99 & 9.6e-04 &     \\
7385478  & 1.65547318   & 0.48 & 1.2e-01 & b   \\
7630658  & 2.15115515   & 0.94 & 2.2e-04 & b   \\
7670617  & 24.70372529  & 0.49 & 1.8e-01 & b   \\
7812175  & 17.79408320  & 0.49 & 1.4e-01 & b   \\
7821010  & 24.23823475  & 0.96 & 6.8e-01 & b   \\
8143170  & 28.78745177  & 0.68 & 1.2e-01 & b   \\
8210721  & 22.67317786  & 0.36 & 1.3e-01 & b   \\
8411947  & 1.79767532   & 0.86 & 1.8e-02 &     \\
8429450  & 2.70515393   & 0.87 & 9.7e-03 & b   \\
8444552  & 1.17809835   & 0.79 & 1.5e-01 & b   \\
8553788  & 1.60616375   & 0.70 & 1.6e-01 & b   \\
8553907  & 42.03215831  & 1.02 & 5.1e-01 & o   \\
8701327  & 8.50883607   & 0.91 & 5.3e-03 &     \\
9007918  & 1.38720670   & 0.48 & 2.8e-04 & b   \\
9053086  & 1.27484170   & 0.55 & 1.5e-01 &     \\
9110346  & 1.79055388   & 0.93 & 9.4e-05 & b   \\
9392702  & 3.90931364   & 0.58 & 1.6e-02 & b   \\
9451096  & 1.25039078   & 0.64 & 2.4e-02 & brc \\
9637299  & 1.88243818   & 0.59 & 3.1e-03 &     \\
9711751  & 1.71152733   & 0.79 & 5.3e-05 & b   \\
9714358  & 6.47419566   & 0.20 & 1.4e-02 & brc \\
9715925  & 6.30827736   & 0.77 & 2.2e-01 & b   \\
9777062  & 19.23003836  & 0.94 & 3.6e-01 &     \\
9850387  & 2.74849856   & 0.93 & 2.8e-03 & b  
\enddata
\vspace{-0.8cm}
\tablecomments{Systems visually identified as exhibiting eclipse timing variations (ETVs). Note the period, mass ratio, and eccentricity included in this table correspond to maximum-likelihood solutions and not the median posterior values. \\
${\dagger}$Triple candidates previously identified by \cite{Borkovits2016}, \cite{Rappaport2013}, \cite{Conroy2014}, and \cite{Orosz2015} contain flags 'b', 'r', 'c', 'o', respectively. 
}
\end{deluxetable}

\begin{deluxetable}{lllll}
\tabletypesize{\footnotesize}
\tablecolumns{5} 
\tablewidth{0pt}
 \tablecaption{ Mass Ratios from APOGEE RVs
 \label{tab:flavien_rvs}}
 \tablehead{
 \colhead{KIC} & \colhead{P (d)} & \colhead{$\Delta$P (d)} & \colhead{Q} & \colhead{$\Delta$Q}  \\ 
 } 
 \startdata  
5284133 & 8.7778    & 0.0178    & 0.5358  & 0.0110    \\
5460835 & 21.47     & 2.64      & 0.8285  & 0.0346    \\
6610219 & 11.300689 & 0.000301  & 0.96134 & 0.00248  
\enddata
\vspace{-0.8cm}
\end{deluxetable}

\begin{deluxetable}{ccccccc}
\tabletypesize{\footnotesize}
\tablecolumns{7} 
\tablewidth{0pt}
 \tablecaption{ \edit{ RV-derived EB Mass Values from Literature}
 \label{tab:ebcompare}}
 \tablehead{
 \colhead{KIC} & \colhead{$P (d)$} & \colhead{$M_1 (M_{\odot})$} & \colhead{ $\Delta M_1 (M_{\odot})$} & \colhead{$M_2 (M_{\odot})$} & \colhead{$\Delta M_2 (M_{\odot})$}  & \colhead{Provenance} 
 } 
 \startdata  

2305372  & 1.40469145  & 1.2               & 0.1                      & 0.62              & 0.04                     & \cite{Matson2017}    \\
3241619  & 1.70334707  & 1.24              & 0.04                     & 0.86              & 0.02                     & \cite{Matson2017}    \\
3440230  & 2.88110031  & 1.6               & 0.1                      & 0.4               & 0.03                     & \cite{Matson2017}    \\
4285087  & 4.4860314   & 1.137             & 0.013                    & 1.103             & 0.014                    & \cite{Clark-Cunningham2019}     \\
4544587  & 2.1891143   & 1.69              & 0.1                      & 1.42              & 0.09                     & \cite{Matson2017}    \\
4574310  & 1.30622004  & 1.38              & 0.06                     & 0.31              & 0.01                     & \cite{Matson2017}    \\
4665989  & 2.248067589 & 1.77              & 0.09                     & 1.32              & 0.05                     & \cite{Matson2017}    \\
4848423  & 3.0036461   & 1.22              & 0.05                     & 1.08              & 0.04                     & \cite{Matson2017}    \\
4851217  & 2.47028283  & 1.43              & 0.05                     & 1.55              & 0.05                     & \cite{Matson2017}    \\
4862625  & 20.000214   & 1.47             & 0.08                    & 0.37             & 0.035                    & \cite{Kostov2013}   \\
5284133  & 8.7845758   & --             & --                    & --             & --                     & F. Kiefer, independent analysis     \\
5285607 & 3.89940111 & 1.557 & 0.038 & 1.346 & 0.033 & \cite{Clark-Cunningham2019} \\
5444392  & 1.51952889  & 1.17              & 0.01                     & 1.19              & 0.1                      & \cite{Matson2017}    \\
5460835  & 21.5392662   & --             & --                    & --             & --                     & F. Kiefer, independent analysis     \\
5473556 & 11.258818 & 1.2207 & 0.0112 & 0.9678 & 0.0039 & \cite{Kostov2016} \\
5513861  & 1.51020825  & 1.5               & 0.04                     & 1.32              & 0.03                     & \cite{Matson2017}    \\
5738698  & 4.80877396  & 1.52              & 0.03                     & 1.44              & 0.02                     & \cite{Matson2017}    \\
5786154  & 2.00827091  & 1.06              & 0.06                     & 1.02              & 0.04                     & \cite{Gaulme2016}    \\
6131659  & 17.5278276  & 0.942            & 0.010                    & 0.703            & 0.008                   & \cite{Clark-Cunningham2019}     \\
6206751  & 1.24534281  & 1.5               & 0.05                     & 0.198             & 0.007                    & \cite{Matson2017}    \\
6525196  & 3.42059774  & 1.0351            & 0.0055                   & 0.9712            & 0.0039                   & \cite{Helminiak2017} \\
6610219  & 21.5392662   & --             & --                    & --             & --                     & F. Kiefer, independent analysis     \\
6762829  & 18.79537    & 0.949             & 0.06                     & 0.2492            & 0.06                     & \cite{Orosz2012b}    \\
6778289  & 30.1301383  & 1.51              & 0.022                    & 1.091             & 0.018                    & \cite{Clark-Cunningham2019}     \\
6781535  & 9.12208635  & 1.01              & 0.03                    & 1.03             & 0.03                    & \cite{Clark-Cunningham2019}     \\
6864859  & 40.8778419  & 1.354             & 0.029                    & 1.411             & 0.028                    & \cite{Clark-Cunningham2019}     \\
7037405  & 207.1083    & 1.25              & 0.04                     & 1.14              & 0.02                     & \cite{Gaulme2016}    \\
7605600 & 3.32619385 & 0.53 & 0.02 & 0.17 & 0.01 
& \cite{Han2019} \\
7821010  & 24.238235   & 1.289             & 0.015                    & 1.231             & 0.015                    & \cite{Helminiak2017} \\
7943602 & 14.69199 & 1.0 & 0.1 & 0.78 & 0.05 & \cite{Gaulme2016} \\
8262223  & 1.61301476  & 1.96              & 0.006                    & 0.21              & 0.001                    & \cite{Guo2017a}      \\
8552540  & 1.06193426  & 1.32              & 0.03                     & 1.04              & 0.02                     & \cite{Matson2017}    \\
8560861  & 31.9732937  & 1.93           & 0.12                   & 1.06            & 0.08                   & \cite{Borkovits2014}     \\
8572936  & 27.7958103  & 1.0479            & 0.0033                   & 1.0208            & 0.0022                   & \cite{Welsh2012}     \\
8823397  & 1.5065037   & 2.1               & 0.2                      & 0.21              & 0.02                     & \cite{Matson2017}    \\
9159301  & 3.04477215  & 1.61              & 0.08                     & 0.4               & 0.02                     & \cite{Matson2017}    \\
9246715  & 171.2768599  & 2.149              & 0.007                     & 2.171               & 0.007                     & \cite{Gaulme2016}    \\
9602595 & 3.5565129 & 3.0 & 0.1 & 0.60 & 0.03 & \cite{Matson2017} \\
9632895  & 27.322037   & 0.934             & 0.01                     & 0.1938            & 0.002                    & \cite{Welsh2014}     \\
9641031  & 2.17815425  & 1.2041            & 0.0076                   & 0.9498            & 0.0046                   & \cite{Helminiak2017} \\
9837578  & 20.733666   & 0.8877            & 0.0053                   & 0.8094            & 0.0045                   & \cite{Welsh2012}     \\
9970396  & 235.2985    & 1.14              & 0.03                     & 1.02              & 0.02                     & \cite{Gaulme2016}   \\
10001167 & 120.3903 & 0.81 & 0.05 & 0.79 & 0.03 & \cite{Gaulme2016} \\
10020423 & 7.44837695  & 1.043             & 0.055                    & 0.362             & 0.013                    & \cite{Orosz2012a}    \\
10031808 & 8.5896432   & 1.741             & 0.009                    & 1.798             & 0.013                    & \cite{Helminiak2017} \\
10156064 & 4.855936446 & 2.1 & 0.1 & 1.44 & 0.08 & \cite{Matson2017} \\
10191056 & 2.427494881 & 1.59              & 0.32                     & 1.427             & 0.036                    & \cite{Helminiak2017} \\
10581918 & 1.80186366 & 1.3 & 0.06 & 0.169 & 0.009 & \cite{Matson2017} \\
10619109 & 2.04516616 & 1.5 & 0.4 & 0.31 & 0.07 & \cite{Matson2017} \\
10736223 & 1.105094186 & 1.6 & 0.1 & 0.35 & 0.03 & \cite{Matson2017} \\
10935310 & 4.128795224 & 0.92 & 0.05 & 0.5 & 0.03 & \cite{Han2017} \\
10987439 & 10.67459809 & 0.9862            & 0.0034                   & 1.4215            & 0.0045                   & \cite{Helminiak2017} \\
11922782 & 3.512934275 & 1.067 & 0.010 & 0.836 & 0.006 & \cite{Helminiak2017} \\
12351927 & 10.116146   & 0.82              & 0.015                    & 0.5423            & 0.008                    & \cite{Kostov2014}    \\
12644769 & 41.07922    & 0.6897            & 0.0035                   & 0.20255           & 0.00066                  & \cite{Doyle2011}     \\
\enddata
\vspace{-0.8cm}
\tablecomments{Note for binary studies which contain multiple mass estimates (e.g., from asteroseismology and RV), we adopted the RV-only derived values for consistency. $\Delta M_1$ and $\Delta M_2$ are uncertainties associated with published mass values. }
\end{deluxetable}

\bsp	
\label{lastpage}
\end{document}